\documentclass{aa}
\usepackage{graphicx}
\usepackage{bm}
\usepackage{txfonts}
\usepackage{natbib}
\usepackage{multirow}
\usepackage{mathrsfs}
\bibpunct{(}{)}{;}{a}{}{,}
\graphicspath{{./fig/}{./png/}}
\newcommand{\uuu}{{\bm u}}

\newcommand{\mBBB}{\overline{\bm B}}

\newcommand{\UUU}{{\bm U}}

\newcommand{\Eq}[1]{Eq.~(\ref{#1})}

\newcommand{\Eqsa}[2]{Eqs.~(\ref{#1}) and (\ref{#2})} 
\newcommand{\Equsa}[2]{Eqs.~(\ref{#1}) to (\ref{#2})}
\newcommand{\EQ}{\begin{equation}}
\newcommand{\EN}{\end{equation}}
\newcommand{\EQA}{\begin{eqnarray}}
\newcommand{\ENA}{\end{eqnarray}}

\newcommand{\pd}{\partial}

\newcommand{\mean}[1]{\overline{#1}}
\newcommand{\meanv}[1]{\overline{\bm #1}}

\newcommand{\cs}{c_{\rm s}}
\newcommand{\cst}{c_{\rm s}^2}

\newcommand{\nut}{\nu_{\rm t}}

\newcommand{\urms}{u_{\rm rms}}

\newcommand{\Beq}{B_{\rm eq}}
\newcommand{\Ma}{{\rm Ma}}

\newcommand{\kf}{k_{\rm f}}

\newcommand{\Co}{{\rm Co}}
\newcommand{\Cost}{\Omega_\star}
\newcommand{\Hp}{H_{\rm p}}

\newcommand{\Pm}{{\rm Pm}}

\newcommand{\Rey}{{\rm Re}}

\newcommand{\Rm}{{\rm Rm}}
\newcommand{\ReM}{{\rm Re}_{\rm M}}

\newcommand{\St}{{\rm St}}
\newcommand{\Lu}{{\rm Lu}}

\newcommand{\Ta}{{\rm Ta}}
\newcommand{\qij}{Q_{ij}}
\newcommand{\qxx}{Q_{xx}}
\newcommand{\qyy}{Q_{yy}}
\newcommand{\qzz}{Q_{zz}}
\newcommand{\qxy}{Q_{xy}}
\newcommand{\qxz}{Q_{xz}}
\newcommand{\qyz}{Q_{yz}}
\newcommand{\tQij}[1]{\tilde{Q}_{#1}}
\newcommand{\qrt}{Q_{r\theta}}
\newcommand{\qrp}{Q_{r\phi}}
\newcommand{\qtp}{Q_{\theta\phi}}

\newcommand{\qpi}{Q_{\phi i}}
\newcommand{\mij}{M_{ij}}
\newcommand{\mpi}{M_{\phi i}}

\newcommand{\tij}{T_{ij}}

{}

\def\onethird{{\textstyle{1\over3}}}

\def\onehalf{{\textstyle{1\over2}}}

\newcommand{\Figa}[1]{Fig.~\ref{#1}}
\newcommand{\Fig}[1]{Figure~\ref{#1}} 
\newcommand{\Figu}[1]{Figure~\ref{#1}}

\newcommand{\Seca}[1]{Sect.~\ref{#1}}
\newcommand{\Sec}[1]{Section~\ref{#1}} 
\newcommand{\Table}[1]{Table~\ref{#1}}

\begin{document}

\authorrunning{K\"apyl\"a}
\titlerunning{Magnetic and rotational quenching of the $\Lambda$ effect}

   \title{Magnetic and rotational quenching of the $\Lambda$ effect}
   \author{P. J. K\"apyl\"a
          \inst{1,2,3,4,5}
          }

   \institute{Georg-August-Universit\"at G\"ottingen, Institut f\"ur Astrophysik,
              Friedrich-Hund-Platz 1, D-37077 G\"ottingen, Germany
              \email{pkaepyl@uni-goettingen.de}
          \and Leibniz-Institut f\"ur Astrophysik, An der Sternwarte 16,
              D-14482 Potsdam, Germany
         \and ReSoLVE Centre of Excellence, Department of Computer Science,
              Aalto University, PO Box 15400, FI-00076 Aalto, Finland
         \and Max-Planck-Institut f\"ur Sonnensystemforschung,
              Justus-von-Liebig-Weg 3, D-37077 G\"ottingen, Germany
         \and NORDITA, KTH Royal Institute of Technology and Stockholm University,
              Roslagstullsbacken 23, SE-10691 Stockholm, Sweden
}


\date{\today}

\abstract
   {%
   Differential rotation in stars is driven by the turbulent transport of
   angular momentum.
   }%
   {
   Our aim is to measure and parameterize the non-diffusive contribution to the total
   (Reynolds plus Maxwell) turbulent stress, known as the $\Lambda$ effect,
   and its quenching as a function of rotation and magnetic field.
   }%
   {
   Simulations of homogeneous, anisotropically forced turbulence in
   fully periodic cubes are used to extract their associated
   turbulent Reynolds and Maxwell
   stresses. The forcing is set up such that the vertical velocity
   component dominates over the horizontal ones,  as in
   turbulent stellar convection. This choice of the forcing
     defines the vertical direction. Additional preferred directions
     are introduced by the imposed rotation and magnetic field
     vectors. The angle between the rotation vector and the
   vertical direction is varied such that the latitude range from the north
   pole to the equator is covered. Magnetic fields are introduced by
   imposing a uniform large-scale field on the system. Turbulent
   transport coefficients pertaining to the $\Lambda$ effect are
   obtained by fitting. The results are compared with analytic studies.
   }%
   {
   The numerical and analytic results agree qualitatively  at
   slow rotation and low Reynolds numbers. This means that vertical
   (horizontal) transport is downward (equatorward). At
   rapid rotation the latitude dependence of the stress is more complex
   than predicted by theory. The existence of a significant meridional
   $\Lambda$ effect is confirmed. Large-scale vorticity generation is
   found at rapid rotation when the Reynolds number exceeds a
   threshold value. The $\Lambda$ effect is severely quenched by
   large-scale magnetic fields due to the tendency of the Reynolds and
   Maxwell stresses to cancel each other. Rotational (magnetic)
   quenching of $\Lambda$ occurs at more rapid rotation (at lower
   field strength) in the simulations than in the analytic studies.
   }%
   {
   The current results largely confirm the earlier theoretical results,
   and  also offer new insights: the non-negligible meridional
   $\Lambda$ effect possibly plays a role in the maintenance of
   meridional circulation in stars, and the appearance of large-scale
   vortices raises the question of their effect on the angular
   momentum transport in rapidly rotating stellar convective
   envelopes. The results regarding magnetic quenching are consistent
   with the strong decrease in differential rotation in recent
   semi-global simulations and highlight the importance of including
   magnetic effects in differential rotation models.
   }%


  \maketitle


\section{Introduction}
\label{sec:intro}

Understanding the causes of solar and stellar differential rotation is
of prime astrophysical interest due to the crucial role that shear
flows play in the generation of large-scale magnetic fields
\citep[e.g.][]{M78,KR80}. The commonly accepted view is that the turbulent
transport of angular momentum is responsible for the generation of
differential rotation in stellar convective envelopes via the
interaction of global rotation and anisotropic turbulence
\citep[e.g.][and references therein]{R89,MT09,RKH13}. The key players
in this respect are the Reynolds and Maxwell stresses that are
correlations of turbulent velocity and magnetic field components,
respectively. Analytic mean-field theories of Reynolds stress driving
of differential rotation in stars have a long history that dates back to
the ideas of \cite{Lebedinski41}, \cite{1946ApNr....4....1W},
\cite{Bi51}, \cite{1963ApJ...137..664K}, and
\cite{1970SoPh...13....3K}; see also the discussion in Chapter~2 of
\cite{R89}.

Much of the current theoretical understanding of solar and stellar
differential rotation is based on the studies of \cite{KR93} and
\cite{KR05} who considered the effects of density stratification and
turbulence anisotropy on turbulent angular momentum transport using
the  second-order correlation approximation (SOCA). They
derived turbulent transport coefficients relevant for the
non-diffusive contribution to the Reynolds stress ($\Lambda$
effect). It can generate differential rotation in a manner similar to the
turbulent $\alpha$ effect, which generates large-scale magnetic fields
\citep[e.g.][]{1966ZNatA..21..369S}.

With an appropriate choice of parameters, two-dimensional axisymmetric
mean-field models that make use of analytic or simplified
descriptions, which in turn are based on heuristic arguments such as
the mixing-length approximation, can reproduce the solar differential
rotation \citep[e.g.][]{Schouea98} with good accuracy
\citep[e.g.][]{BMT92,KR05,Re05,2016AdSpR..58.1490P,2017ApJ...835....9B}.
Furthermore, studies of other late-type stars and the dependence on
the rotation rate and spectral type
\citep[][]{2005A&A...433.1023K,2005AN....326..265K,2011A&A...530A..48K,KO11,2012MNRAS.423.3344K}
are in reasonable agreement with observations
\citep[e.g.][]{1991LNP...380..353H,Henry1995,RRB13}.

Most of the these studies neglect magnetic fields, which are
very likely to have a significant impact on turbulent transport in
real stars. The influence of large-scale magnetic fields on the
$\Lambda$ effect and the consequences of the ensuing magnetic
quenching of differential rotation have been studied by several
authors in the mean-field framework
\citep{KRK94,1996A&A...312..615K,2017MNRAS.466.3007P,2018JASTP.179..185P}.
An additional complication arises from the fact that turbulent
viscosity also depends on rotation and large-scale magnetic fields
\citep[e.g.][]{R89,KPR94,Ki16}. Arguably, however, the biggest caveat of the
analytic studies is that current approaches cannot be used
to derive turbulent transport coefficients rigorously beyond the
validity range of SOCA which requires that ${\rm min}(ul/\nu,\tau_{\rm
  c}u/l)={\rm min}(\Rey,\St)$, where $u$ and $l$ are the typical
velocity and length scale, $\tau_{\rm c}$ is the correlation time of
the turbulence, and $\Rey$ and $\St$ are the Reynolds and Strouhal
numbers, respectively \citep[e.g.][]{R80,KR80}. The constraint on the
Reynolds number is far removed from the parameter regimes of real
astrophysical objects \citep[e.g.][]{BS05}. Although numerical
simulations fall short in comparison to real astrophysical systems in
terms of the Reynolds number, they still typically have
$\Rey\gg1$. Furthermore, estimates of the Strouhal number from
simulations suggest that $\St\approx1$
\cite[e.g.][]{2004PhFl...16.1020B,2005A&A...439..835B}.

Extracting turbulent transport coefficients from numerical simulations
appears to be an obvious remedy to the restrictions of SOCA. Several
authors have used three-dimensional simulations of turbulent
convection in local Cartesian
\citep[e.g.][]{1984ApJ...276..316H,PTBNS93,1998ApJ...493..955B,Chan01,KKT04,2005AN....326..315R}
and in global and semi-global set-ups in spherical coordinates to
compute the turbulent angular momentum transport
\cite[e.g.][]{RBMD94,KMGBC11,GSKM13,KKB14,HRY15a,WKKB16}. However, a unique
separation of the contributions from different effects is currently
not possible. This is already apparent  from a truncated expression of
the Reynolds stress in rotating and anisotropic turbulence
\begin{eqnarray}
\qij=\Lambda_{ijk} \Omega_k - \mathcal{N}_{ijkl} \mean{U}_{k,l}+\ldots,
\label{equ:Qij}
\end{eqnarray}
where $\qij=\mean{u_i u_j}$ is the Reynolds stress (the velocity
fluctuations $u_i=U_i-\mean{U}_i$ are the differences between the
total ($U_i$) and mean ($\mean{U}_i$) velocities) and $\Omega_i$
is the rotation vector. Furthermore, the overbar denotes suitable
averaging, and the dots on the right-hand side indicate the
possibility of terms proportional to higher-order derivatives. The
coefficients $\Lambda_{ijk}$ and $\mathcal{N}_{ijkl}$ are third- and
fourth-rank tensors, respectively, and describe the $\Lambda$ effect
and turbulent viscosity. It is immediately clear that in the general
case, the right-hand side of \Eq{equ:Qij} contains more unknowns
than the simulations provide in the form of $\qij$.

A way to circumvent the lack of constraints on \Eq{equ:Qij} is to make
simplifying assumptions, for example employing a turbulent viscosity
to compute the components of $\Lambda$
\cite[e.g.][]{KBKSN10,KKB14,KKKBOP15,WKKB16}. Another possibility is
to perform fitting that, for example, involves truncating
\Eq{equ:Qij} at the desired order and forming a sufficient number of
moments with other large-scale quantities such that the number of
equations and the components of $\Lambda_{ijk}$ and
  $\mathcal{N}_{ijkl}$ are equal (e.g.\, \cite{2018A&A...611A..15K};
see
also \cite{2002GApFD..96..319B}). The coefficients from the truncated
\Eq{equ:Qij} are then obtained by inverting a simple matrix. This
procedure effectively uses the turbulent transport coefficients as fit
parameters. However, the accuracy of these methods has not been
thoroughly studied. In mean-field electrodynamics a similar issue
appears in conjunction with the electromotive force and its expansion
in terms of large-scale magnetic fields and their gradients
\citep[e.g.][]{KR80}. There the situation is significantly simpler in
that a linear relation between the electromotive force and the mean
magnetic field exists in the kinematic regime where the field is weak.
In this case it is possible to extract all relevant turbulent
transport coefficients by solving a sufficient amount of independent
problems with imposed test fields within the framework of the
test-field method
\cite[e.g.][]{SRSRC05,SRSRC07}. Furthermore, the method has been shown
to produce results consistent with mean-field theory even in the
non-kinematic regime
\citep{2008ApJ...687L..49B,2010A&A...520A..28R}. Formulating a similar
procedure for the Navier--Stokes equations is, however, much more
challenging due to their inherent non-linearity.

Currently the most reliable option for extracting turbulent transport
coefficients relevant for angular momentum transport is to reduce the
system to the fewest  possible ingredients in order to minimize the
ambiguity in the interpretation. This approach was adopted in an
earlier work by \cite{KB08} who examined the $\Lambda$ effect in
anisotropically forced homogeneous turbulence under the influence of
rotation in triply periodic cubes. In the current study the same model
is used and the primary interest is to further refine the picture of
the $\Lambda$ effect in the case where the minimum requirements for
its existence are present.  Here the rotational quenching and Reynolds
number dependence of the $\Lambda$ effect are studied in detail with
significantly extended coverage of the respective parameter ranges in
comparison to \cite{KB08}. Furthermore, motivated by recent global and
semi-global convection simulations where the differential rotation is
severely affected by magnetic fields
\citep{2016AdSpR..58.1507V,2017A&A...599A...4K}, the task of measuring
the quenching of the $\Lambda$ effect as a function of an imposed
large-scale magnetic fields is undertaken for the first time. The
numerical results are compared in detail with analytic SOCA
expressions derived by \cite{KR05} and \cite{KRK94}. The current study
is a step toward building a mean-field framework of differential
rotation modelling  in parameter regimes no longer restricted by the
limitations of SOCA.

\section{Turbulent angular momentum transport and $\Lambda$ effect}

\subsection{Theoretical considerations}

The azimuthally averaged angular momentum in the $z$ direction in
spherical polar coordinates is governed by the equation
\begin{eqnarray}
&& \frac{\pd}{\pd t} (\mean{\rho} \varpi^2 \Omega)
+ \bm\nabla\bm\cdot\{\varpi [\varpi\mean{\rho \bm{U}}\Omega
+ \mean{\rho}\ \qpi
- 2\nu\mean{\rho} \mean{\bm{\mathsf{S}}}\bm\cdot{\hat{\bm\phi}}
\nonumber \\ && \hspace{4cm}
-\mu_0^{-1}\mean{B}_\phi\meanv{B} - \mpi] \}=0,
\label{equ:angmom}
\end{eqnarray}
where $\rho$ is the density, $\varpi=r\sin\theta$ is the lever arm,
$\Omega$ is the angular velocity, $\nu$ is the kinematic viscosity,
$\bm{\mathsf{S}}$ is the rate of strain tensor defined below, and
$\hat{\bm{\phi}}$ is a unit vector in the azimuthal direction. The
velocity and the magnetic field have been decomposed into their mean
and fluctuating parts, such that $U_i =\mean{U}_i + u_i$ and $B_i
=\mean{B}_i + b_i$, respectively. In addition to the Reynolds stress
defined above, \Eq{equ:angmom} includes the Maxwell stress
$\mij=\mu_0^{-1}\mean{b_i b_j}$. Here the correlations of density
fluctuations with the velocity have been omitted, which is valid for
incompressible flows. In the current study the flows are weakly
compressible with Mach number  of the order of 0.1, and so  this
omission should only play  a minor role.

In stars, turbulent Reynolds and Maxwell stresses are the main
contributors to the angular momentum transport
\citep[e.g.][]{MT09,RKH13}. Much of mean-field theory deals with the
task of representing the turbulent correlations in terms of
large-scale quantities,  as in \Eq{equ:Qij}.  Here the relation
between the turbulent stress and the large-scale quantities is assumed to be
local and instantaneous. Spatial and temporal non-locality are in
general non-negligible in turbulent flows
\citep[e.g.][]{BRS08,2009ApJ...706..712H}. However, dealing with this
generalization will be saved for a future study. The first term on the
 right-hand side of \Eq{equ:Qij} describes the $\Lambda$ effect, or the
non-diffusive contribution, to the Reynolds stress in rotating
anisotropic turbulence. In addition, a term proportional to
large-scale velocities ($\Gamma_{ijk}\mean{U}_k$) can also appear on
the right-hand side of \Eq{equ:Qij} (see \citealt{FSS87}). This
corresponds to the anisotropic kinetic alpha (AKA) effect which arises
in non-Galilean invariant flows
\citep[e.g.][]{FSS87,BvR01,2018A&A...611A..15K}. The forcing used in
the present study is Galilean invariant, and thus the contributions
from the AKA effect are likely to be negligible.  Even so, the current
data analysis method cannot detect the AKA effect.

In spherical coordinates the Reynolds stress components
$\qrp=\mean{u_r u_\phi}$ and $\qtp=\mean{u_\theta u_\phi}$ enter the
angular momentum equation directly and correspond to radial and
latitudinal fluxes of angular momentum. The third off-diagonal
component, $\qrt=\mean{u_r u_\theta}$, contributes to the maintenance
of meridional flows and thus influences the angular momentum balance
indirectly \citep[e.g.][]{R89}. The non-diffusive part of this stress
component, the `meridional' $\Lambda$ effect, is typically omitted
in models of solar and stellar differential rotation. However,
numerical simulations indicate that this contribution is
non-negligible \citep{PTBNS93,RBMD94,KB08}, and that it may be
important in driving the meridional flows in the near-surface layers
of the Sun \citep{HRY15a,WKKB16}. When approximated with an
  isotropic and homogeneous (constant) turbulent
viscosity, the corresponding components of the $\Lambda$ effect can be
formulated as
\begin{eqnarray}
\qrp^{(\Lambda)} &=& \nut \Omega \mathscr{V}, \label{equ:Qrp} \\
\qtp^{(\Lambda)} &=& \nut \Omega \mathscr{H}, \label{equ:Qtp} \\
\qrt^{(\Lambda)} &=& \nut \Omega \mathscr{M}, \label{equ:Qrt}
\end{eqnarray}
where $\nut$ is the turbulent viscosity, $\mathscr{V} = V \sin
\theta$, $\mathscr{H} = H \cos \theta$, and $\mathscr{M} = M \sin
\theta \cos \theta$. The factor $\nut \Omega$ is used for
normalization, whereas $V$, $H$, and $M$ are dimensionless. These three factors
are often expanded in powers of $\sin^2\theta$ \citep[e.g.][]{BTMR90}
\begin{eqnarray}
  V &=& V^{(0)} + V^{(1)} \sin^2 \theta + V^{(2)} \sin^4 \theta + \ldots, \label{equ:V} \\
  H &=& H^{(0)} + H^{(1)} \sin^2 \theta + H^{(2)} \sin^4 \theta + \ldots, \label{equ:H} \\
  M &=& M^{(0)} + M^{(1)} \sin^2 \theta + M^{(2)} \sin^4 \theta + \ldots, \label{equ:M}
\end{eqnarray}
where the dots indicate the possibility of higher-order
terms\footnote{Sometimes $V$ is also expanded  in terms of powers of $\cos^2
  \theta$ \citep[e.g.][]{KR05}.}. We  also note that in general the
  coefficients are functions of position, that is $V=V({\bm x})$,
  $V^{(i)}=V^{(i)}({\bm x})$, and so on. These coefficients can be
written more
compactly as
\begin{eqnarray}
  \mathscr{V}^{(j)} &=& \left(\sum_{i=0}^{j} V^{(i)} \sin^{2i} \theta \right)\sin\theta, \label{equ:VV} \\
  \mathscr{H}^{(j)} &=& \left(\sum_{i=0}^{j} H^{(i)} \sin^{2i} \theta \right)\cos\theta, \label{equ:HH} \\
  \mathscr{M}^{(j)} &=& \left(\sum_{i=0}^{j} M^{(i)} \sin^{2i} \theta \right)\sin\theta \cos\theta. \label{equ:MM}
\end{eqnarray}
In stellar convection zones, buoyancy drives the convective
  instability, which can be considered  a large-scale anisotropic
  forcing of turbulence \citep[e.g.][]{1992PhRvL..69..769Y}. This
forcing is expected to lead to turbulence
dominated by the radial velocity \citep[e.g.][]{KMB11}. With 
turbulence of this kind in the slow rotation limit, only the vertical $\Lambda$
effect is expected to survive. More specifically, the `fundamental'
mode of the vertical $\Lambda$ effect, with the corresponding
coefficient $V^{(0)}$, is expected to tend to a constant value, that
is $V^{(0)}\rightarrow \mbox{const}$ for $\Omega\rightarrow0$
\cite[e.g.][]{R89,KR93}. A corresponding horizontal effect, described
by $H^{(0)}$, does not arise in the hydrodynamic case because it would
require another preferred direction to be present in the system
\citep{R89}. Large-scale magnetic fields can provide this additional
anisotropy in which case $H^{(0)}\neq0$ \citep{KRK94}. The higher-order components of the vertical and horizontal $\Lambda$ coefficients
are expected to be proportional to higher (even) powers of $\Omega$
due to the symmetry properties of $\Lambda_{ijk}$
\citep{R89,RKT14}. The theory of the meridional $\Lambda$ effect is
much less well developed\footnote{See, however, p.\ 117 of \cite{R89}
  and \cite{1989A&A...217..217T}.}.

\subsection{Observational evidence}

Traditionally the most often used argument in favour of the existence
of the $\Lambda$ effect has been the observed cross-correlation of
horizontal proper motions of sunspots, which have been interpreted as
the $\qtp$ component of the Reynolds stress
\citep[e.g.][]{1965ApJ...141..534W,1998A&A...332..755P}. These studies
yield a value of the order of $10^3$~m$^2$~s$^{-1}$ which is positive
(negative) in the northern (southern) hemisphere of the Sun. If the
Reynolds stress is assumed to result from the shear stress alone
(Boussinesq ansatz),
\begin{equation}
\qij = - \nut \left(\mean{U}_{i,j} + \mean{U}_{j,i}\right), \label{equ:Boussinesq}
\end{equation}
the $\qtp$ component in spherical coordinates is
\begin{equation}
\qtp = - \nut \sin \theta \frac{\pd \Omega}{\pd \theta}.\label{equ:qtpBou}
\end{equation}
This implies that $\nut < 0$ given that $\sin \theta >0$ and
$\frac{\pd \Omega}{\pd \theta} > 0$ in the northern hemisphere of the
Sun \citep[see the discussions in][]{1989A&A...217..217T,PTBNS93}. A
more recent result from giant cells, obtained from supergranulation
tracking, suggests the same sign but two orders of magnitude lower
amplitude for $\qtp$ \citep{HUC13}. Although a negative $\nut$ has
been reported to occur in certain cases of two-dimensional turbulence
\citep[e.g.][]{KR74b}, there is no evidence for $\nut<0$ for more
general three-dimensional flows. Negative turbulent viscosity would
also lead to a pile-up of energy at small scales which is considered
unphysical. A more plausible explanation is that \Eq{equ:qtpBou} is
incomplete and that a non-diffusive term ($\Lambda$ effect) has to be
present. In this  case, a positive $\qtp$ in the northern hemisphere
can occur when the rotational influence on the flow is sufficiently
large, such that the contribution of the $\Lambda$ effect to the
Reynolds stress dominates over that from turbulent viscosity
\citep{RKT14}. Thus, sunspots and giant cells are likely to probe these
deeper, and more rotationally constrained, parts of the convection
zone. However, due to the vanishing horizontal $\Lambda$ effect near
the surface \citep{RKT14}, the viscous term, \Eq{equ:Boussinesq}, is
expected to dominate there and to produce a negative $\qtp$. This
detection from helioseismology of supergranulation has been reported
recently by \cite{2016AnRFM..48..191H}.

On the other hand, near the solar surface, where the rotational
influence is weak \citep[e.g.][]{GHFT15,GHT16a}, only the fundamental
mode of the $\Lambda$ effect is expected to be non-zero. This
coincides with the near-surface shear layer where the angular velocity
has a uniform radial gradient of $\Omega$ independent of latitude
\citep{2014A&A...570L..12B}. This can be explained with a simple model
where the meridional flows are neglected and where only the
fundamental mode of the $\Lambda$ effect is taken into account
\citep[e.g.][]{1982ApL....22...89G,Ki16}, yielding
\begin{equation}
\frac{\pd \ln \Omega}{\pd \ln r} = V^{(0)} \approx -1,
\end{equation}
where the latter is the observational result of
\cite{2014A&A...570L..12B}. No direct observational data for the
behaviour  of $\qrp$ in the deep parts or of $\qrt$ anywhere in the
solar convection zone are currently available. Using a combination of
theoretical arguments and helioseismic measurements,
\cite{2011ApJ...743...79M} infer that the radial angular momentum
transport attributed to turbulent convection in the near-surface shear
layer is radially inward. The Boussinesq ansatz, $\qrp = - \nut r
\sin\theta \frac{\pd \Omega}{\pd r}$, would again imply $\nut<0$ in
the absence of a radial $\Lambda$ effect.

\section{Model} \label{sect:model}

Compressible turbulent fluid flow is modelled in a triply periodic cube
with or without magnetic fields. The gas is assumed to obey an
isothermal equation of state $p=\rho \cst$, where $p$ is the pressure
and $\cs$ is the constant speed of sound. Gravity is neglected for
simplicity as it would necessarily introduce inhomogeneity in a
compressible system, and because the minimal requirements for the
appearance of the $\Lambda$ effect can be achieved by a
homogeneous
forcing. The system is governed by the induction, continuity, and
Navier--Stokes equations
\begin{eqnarray}
  \frac{\pd {\bm A}}{\pd t} &=& {\bm U}\times [ {\bm B} + {\bm B}^{(0)}] - \eta\mu_0{\bm J}, \\
  \frac{D \ln \rho}{Dt} &=& -\bm\nabla\bm\cdot{\bm U}, \\
  \frac{D {\bm U}}{Dt} &=& -\cst \bm\nabla\ln\rho - 2\ \bm\Omega \times {\bm U} + {\bm F}^{\rm visc} + {\bm F}^{\rm force},
\end{eqnarray}
where ${\bm A}$ is the magnetic vector potential, ${\bm U}$ is the
velocity, ${\bm B} = \bm\nabla \times {\bm A}$ is the magnetic field,
${\bm B}^{(0)}$ is the uniform imposed external magnetic field,
$\eta$ is the magnetic diffusivity, $\mu_0$ is the permeability of
vacuum, ${\bm J} = \mu_0^{-1} \bm\nabla \times {\bm B}$ is the current
density, $D/Dt = \pd/\pd t -{\bm U} \bm{\cdot\nabla}$ is the
advective time derivative, $\rho$ is the density, $\bm\Omega$ is the
rotation vector, and ${\bm F}^{\rm visc}$ and ${\bm F}^{\rm
  force}$ respectively describe the viscous force and external forcing.

The viscous force reads
\begin{eqnarray}
{\bm F}^{\rm visc} = \nu \left(\nabla^2\UUU + \onethird \bm\nabla\bm\nabla\bm\cdot \UUU + 2\bm{\mathsf{S}} \bm\cdot \bm\nabla\ln\rho \right),
\end{eqnarray}
where $\nu$ is the constant kinematic viscosity and $\mathsf{S}_{ij} = \onehalf
(U_{i,j}+ U_{j,i}) - \onethird \delta_{ij} U_{k,k}$ is the traceless
rate of strain tensor where the commas denote differentiation.

The external forcing is given by
\begin{eqnarray}
{\bm F}^{\rm force}({\bm x},t) = Re\{\bm{\mathsf{N}}\bm\cdot{\bm f}_{{\bm k}(t)} \exp[{\rm i} {\bm k}(t)\bm\cdot {\bm x} - {\rm i}\phi(t)]\},
\end{eqnarray}
where ${\bm x}$ is the position vector and $\bm{\mathsf{N}}= \bm{\mathsf{f}}\cs
(k\cs/\delta t)^{1/2}$ is a tensorial normalization factor. Here $\bm{\mathsf{f}}$
contains the non-dimensional amplitudes of the forcing (see below), $k=|{\bm
  k}|$, $\delta t$ is the length of the time step, and $-\pi < \phi(t) < \pi$
is a random delta-correlated phase. The vector ${\bm f}_{\bm k}$ describes
non-helical transversal waves, and is given by
\begin{eqnarray}
{\bm f}_{\bm k} = \frac{{\bm k} \times \hat{\bm e}}{\sqrt{ {\bm k}^2 - ({\bm k} \bm\cdot \hat{\bm e})^2 }},
\end{eqnarray}
where $\hat{\bm e}$ is an arbitrary unit vector, and where the
wavenumber ${\bm k}$ is randomly chosen. The {\sc Pencil
  Code}\footnote{\url{http://github.com/pencil-code}} was used to
produce the numerical simulations.

\subsection{Units and system parameters}

The units of length, time, density, and magnetic field are
\begin{eqnarray}
[x] = k_1^{-1}, [t]=(\cs k_1)^{-1}, [\rho] = \rho_0, [B] = \sqrt{\mu_0\rho_0} c_s,
\end{eqnarray}
where $\rho_0$ is the initially uniform value of density.
The forcing amplitude is given by
\begin{eqnarray}
\mathsf{f}_{ij} = f_0 (\delta_{ij} + \delta_{iz} \cos^2
\Theta_{\bm k} f_1/f_0)
,\end{eqnarray}
where $f_0$ and $f_1$ are the amplitudes of the isotropic and
anisotropic parts, respectively. Furthermore, $\delta_{ij}$ is the
Kronecker delta, and $\Theta_{\bm k}$ is the angle between the
vertical direction and ${\bm k}$. The forcing wavenumber is chosen
from a narrow range $9.9 \leq \kf/k_1 \leq 10.1$. The set of
wavenumbers consists of 318 unique combinations of
$\bm{k}=(k_x,k_y,k_z)$, which are uniformly distributed \citep[for
  more details about the forcing, see][]{B01}. The amplitude of the
forcing is chosen such that the Mach number, $\Ma=\urms/\cs$, where
$\urms$ is the volume-averaged rms velocity, is of the order of 0.1.

The remaining system parameters in the hydrodynamic cases are the
kinematic viscosity $\nu$, and the rotation vector
$\bm\Omega=\Omega_0(-\sin\theta,0,\cos\theta)^{\rm T}$, where $\theta$
is the angle that the rotation vector makes with the vertical ($z$)
direction. An angle of $\theta=0$ ($90\degr$) corresponds to the
north pole
(equator). Viscosity and rotation can be combined into the Taylor
number
\begin{eqnarray}
\Ta = \frac{4\,\Omega_0^2 L_{\rm d}^4}{\nu^2},
\end{eqnarray}
where $L_{\rm d}=2\pi/k_1$ corresponds to the size of the computational domain and
where $k_1$ is the wavenumber corresponding to the box size.

In magnetohydrodynamic (MHD) cases the magnetic diffusivity is an additional
control parameter. This is quantified by the magnetic Prandtl number, which is
the ratio of the kinematic viscosity and magnetic diffusivity:
\begin{eqnarray}
\Pm = \frac{\nu}{\eta}.
\end{eqnarray}
All of the simulations in the present study use $\Pm=1$.
The imposed magnetic field strength is measured by the Lundquist number
\begin{eqnarray}
\Lu = \frac{v_{\rm A}}{\eta \kf},
\end{eqnarray}
where $v_{\rm A}=B_0/(\mu_0 \rho_0)$ is the Alfv\'en speed and $B_0$
is the amplitude of the imposed field. The choice of a purely
toroidal imposed field comes from turbulent mean-field models of the
solar dynamo \citep[e.g.][]{KKT06,2013ApJ...776...36P} and
three-dimensional global simulations which suggest that the strong
differential rotation produces a dominant toroidal field in the bulk
of the convection zone \citep[e.g.][]{GCS10,KMB12,NBBMT13}.

\subsection{Diagnostics quantities}

The following quantities are outcomes of the simulations that can only
be determined a posteriori. The fluid and magnetic Reynolds numbers are
given by
\begin{eqnarray}
\Rey = \frac{\urms}{\nu \kf},\ \ \  \ReM = \frac{\urms}{\eta \kf}.
\end{eqnarray}
The rotational influence on the flow is quantified by the Coriolis
number based on the forcing scale
\begin{eqnarray}
\Omega_\star = \frac{2\,\Omega_0 \ell}{\urms},\label{equ:Co}
\end{eqnarray}
where $\ell=L_{\rm d}k_1/\kf=2\pi/\kf$. The Strouhal number is given
by
\begin{eqnarray}
\St = \frac{\urms}{\ell} \tau,\label{equ:St}
\end{eqnarray}
where $\tau$ is the correlation time of the flow.

The magnetic field strength is often quoted in terms of the
equipartition value
\begin{eqnarray}
\Beq=(\mu_0 \rho \UUU^2)^{1/2}.
\end{eqnarray}
Finally, the parameters
\begin{eqnarray}
  A_{\rm V} = \frac{\qxx+\qyy-2\qzz}{\urms^2},\ \ \ A_{\rm H} = \frac{\qyy-\qxx}{\urms^2}.
\end{eqnarray}
characterize the vertical and horizontal anisotropy of turbulence.

\begin{figure*}
  \centering
  \includegraphics[width=0.9\textwidth]{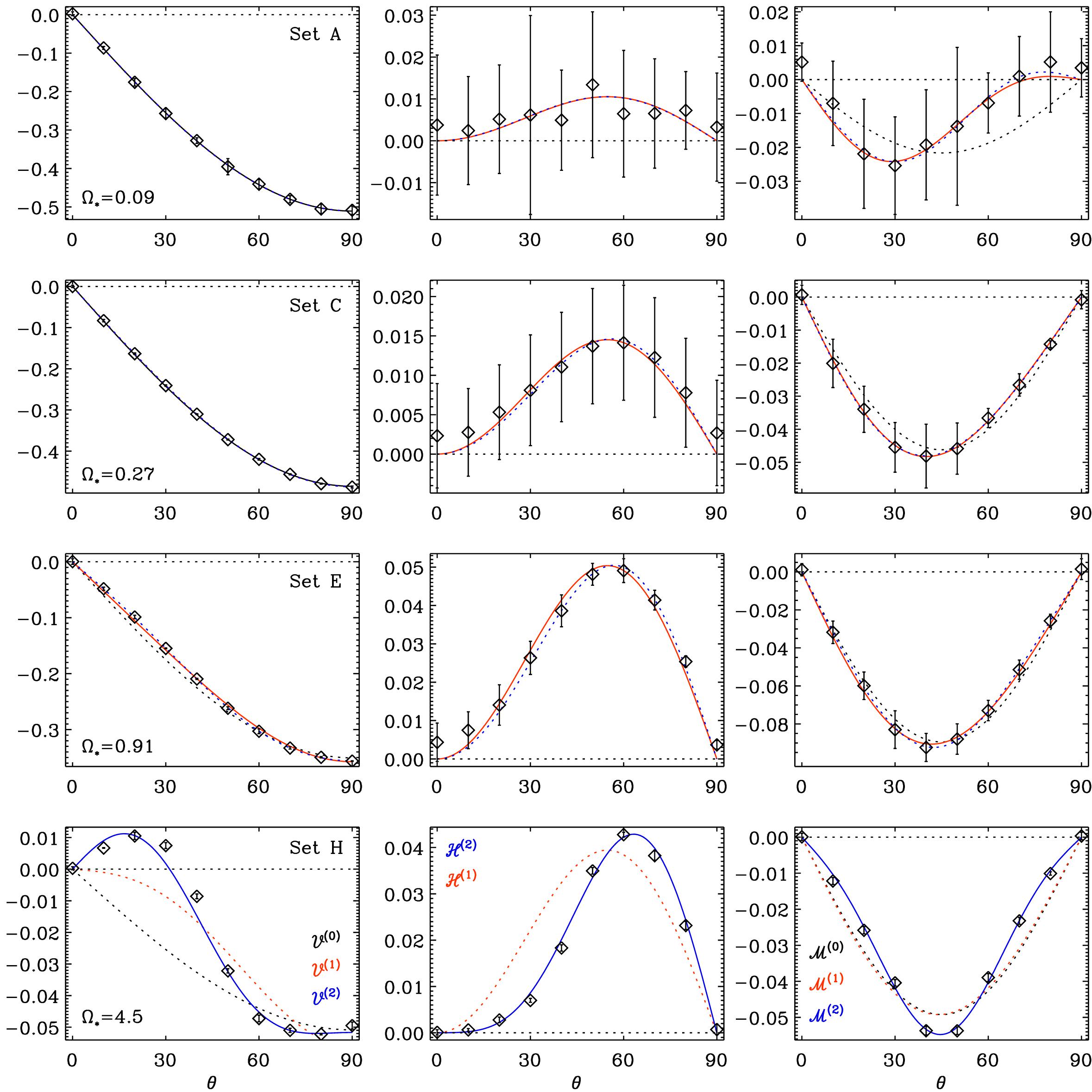}
\caption{From left to right: $\mathscr{V}$, $\mathscr{H}$, and
  $\mathscr{M}$ from \Equsa{equ:Vsim}{equ:Msim}, along with the fits
  to \Equsa{equ:VV}{equ:MM} where one (black), two (red), or three
  (blue) first terms of the expansions are retained. Data from Sets~A
  (top row), C, E, and H (bottom).}
\label{fig:pstress}
\end{figure*}

\subsection{Modelling and data analysis strategies}
\label{sec:strategy}

Numerous sets of simulations were performed where a single physical
ingredient (rotation, imposed magnetic field, viscosity) was varied
and  the other system parameters were kept fixed. In most cases
each set of simulations consisted of ten runs in which the colatitude
$\theta$ was varied in steps of $10\degr$ from the north pole to the
equator. The only exceptions to this are Sets~AI1--AI4 and AI1h, which
were done at a fixed colatitude of $\theta=45\degr$. The simulations
are summarized  in Tables~\ref{tab:runs_aniso}--\ref{tab:runs_mag}.
The grid resolution of the bulk of the simulations is low ($144^3$) in
order to cover large parameter ranges with a reasonable computational
cost. Higher resolutions were used  in cases with a larger
scale-separation ratio (Set AI1h) and in those cases with a higher
Reynolds number (Sets RE7 and RE8).

The simulations were made in the $f$--plane approximation in
Cartesian coordinates. For the off-diagonal Reynolds stresses this
means moving from spherical polar coordinates to local Cartesian ones,
implying $(r,\theta,\phi) \rightarrow (z,x,y)$. Combining this with
\Equsa{equ:Qrp}{equ:MM} and \Eq{equ:Co}, and using the estimate
$\nut={4 \over 15} \urms \ell$ for the turbulent viscosity
\citep{KPR94}, yields
\begin{eqnarray}
\mathscr{V} &=& \tQij{yz} \mathscr{R}, \label{equ:Vsim} \\
\mathscr{H} &=& \tQij{xy} \mathscr{R}, \label{equ:Hsim} \\
\mathscr{M} &=& \tQij{xz} \mathscr{R}, \label{equ:Msim}
\end{eqnarray}
where the tilde refers to normalization by $\urms^2$, and where
\begin{eqnarray}
\mathscr{R} = \frac{15}{2}\Omega_\star^{-1}.\label{equ:fancyR}
\end{eqnarray}
Furthermore, in the local domain approximation  in the absence of
gravity the coefficients $\mathscr{V}$, $\mathscr{H}$, and $\mathscr{M}$
are independent of position. It is useful to parameterize the
Reynolds stress in terms of the coefficients \Equsa{equ:V}{equ:M} to
make the results more readily comparable with theory and usable in
mean-field modelling. The numerical data from
\Equsa{equ:Vsim}{equ:Msim} is thus fitted with \Equsa{equ:VV}{equ:MM}
with varying number of terms included. The issues related to fitting
(\Seca{sec:intro}) are alleviated by the fact that no large scale
flows are present in the current simulations, and the
off-diagonal components of the Reynolds stress can be considered to
arise solely due to the $\Lambda$ effect. However, the choice of the
$\theta$-dependence of the fit for the coefficients $V^{(i)}$,
$H^{(i)}$, and $M^{(i)}$ still leads to non-uniqueness of the
extracted coefficients.

The following procedure was used in the present study: the
coefficients $V^{(i)}$, $H^{(i)}$, and $M^{(i)}$ were obtained from
fits to \Equsa{equ:VV}{equ:MM} for each set of simulations
corresponding to a fixed Taylor number. Three criteria were considered
in evaluating the adequacy of the fit:
\begin{enumerate}
\item The fit is considered acceptable if the rms deviation of the fit
  from the numerical data,
  \begin{equation}
    \delta\qij^{\rm fit}=\sqrt{\frac{1}{N+1}\sum_{k=1}^N \left[\qij^{\rm fit}(\theta_k) -
    \qij(\theta_k) \right]^2},
  \end{equation}
  is smaller than the rms error of the data
  \begin{equation}
    \delta\qij^{\rm data}=\sqrt{\frac{1}{N+1}\sum_{k=1}^N \left[\delta\qij^{\rm data}(\theta_k)\right]^2},
  \end{equation}
  where $N+1=10$ is the number of data points in each set,
  $\theta_k=k\cdot 10\degr$, and where $\delta\qij^{\rm
    data}(\theta_k)$ are the errors of the individual data points. The
  latter were computed by dividing the time series in three parts and
  averaging over each part. The greatest deviation of these from the
  average over the full data set was taken to represent the error.
\item Independent from criterion 1, an additional constraint is
  imposed on $\delta\qij^{\rm fit}$: the expansions in
  \Equsa{equ:VV}{equ:MM} are truncated if an additional term does not
  decrease $\delta\qij^{\rm fit}$ by more than $10$ \%.
\item In some cases, when the latitudinal profile of the stress
  changes as a function of $\Omega_\star$, equally good fits may be
  found for a higher- and lower-order expansion. In these cases, the
  higher-order term was taken into account if it made the
  $\Omega_\star$ dependence of the coefficient smoother.
\end{enumerate}
In practice, Criterion 1 cannot be fulfilled on many occasions. In
these cases, Criterion 2 becomes decisive, as is demonstrated in the
left and middle panels on the lower row of
\Figa{fig:pstress_Co}. Criterion 3 was invoked twice: in Set~E the
fits with $\mathscr{H}^{(1)}$ and $\mathscr{H}^{(2)}$, and
$\mathscr{M}^{(1)}$ and $\mathscr{M}^{(2)}$, respectively, were
roughly equally good but the latter choices lead to a smoother
rotational dependence of $H^{(i)}$ and $M^{(i)}$ (see
\Seca{subsec:latdep}).

\begin{table*}
\centering
\caption[]{Summary of runs with varying rotation and turbulence
  anisotropy at $\theta=45\degr$.}
  \label{tab:runs_aniso}
       \vspace{-0.5cm}
      $$
          \begin{array}{p{0.05\linewidth}ccccccccccccc}
          \hline
          \hline
          \noalign{\smallskip}
          Set & \Ta\ [10^6] & f_0\ [10^{-3}] & f_1/f_0 & A_{\rm V} & A_{\rm H}\ [10^{-4}] & \Omega_\star \\
          \hline
          AI1  & 0.156\ldots15.6 &     10^{-3}     & 4\cdot 10^4 & -0.50 \ldots -0.52 & 1 \ldots 40 & 0.05\ldots0.46 \\ %
          AI1h & \ \ \ 6.93\ldots1560  &     10^{-3}     & 4\cdot 10^4 & -0.56 \ldots -0.58 & 2 \ldots 42 & 0.03\ldots0.47 \\ %
          AI2  & 0.156\ldots15.6 &     3.0  &       9     & -0.38 \ldots -0.39 & 2 \ldots 30 & 0.05\ldots0.46 \\ %
          AI3  & 0.156\ldots15.6 &     3.0  &       4     & -0.24 \ldots -0.25 & 3 \ldots 18 & 0.05\ldots0.46 \\ %
          AI4  & 0.156\ldots15.6 &     4.5 &      1.2    & -0.08 & \!\! 2 \ldots 8 & 0.05\ldots0.45 \\ %
          \hline
          \end{array}
          $$ \tablefoot{Sets~AI1--4 have $\tilde{k}_{\rm f}=10$ and
            Set~AI1h $\tilde{k}_{\rm f}=30$, where the tildes refer to
            normalization by $k_1$. The Reynolds number is 14 in
            Sets~AI1--4 and 15 in Set~AI1h. Grid resolutions $144^3$
            (Sets~AI1--4) and $288^3$ (AI1h) were used.}
\end{table*}

Once a higher-order term in the expansion is included, it is retained
for all subsequent sets at higher Taylor numbers. Examples of the
different fits are shown for a representative selection of runs in
\Fig{fig:pstress}, where the solid lines indicate the accepted fit
according to the procedure outlined above and the dotted lines
indicate discarded fits.

\section{Results} \label{sect:results}

\subsection{Reynolds stress in the slow rotation limit}

Quasi-linear theory of the $\Lambda$ effect predicts that for
vertically dominated turbulence only the  fundamental
mode of the vertical $\Lambda$ effect, corresponding to $V^{(0)}$ in
\Eq{equ:V}, is present in the limit of slow rotation
\mbox{\citep{R89,KR93}}. This implies that vertical Reynolds stress is
linearly proportional to $\Omega$ in this regime, or
\begin{eqnarray}
\tQij{yz}^{(\Lambda)} = {2 \over 15} \Omega_\star V^{(0)} \sin \theta.
\end{eqnarray}
The other off-diagonal components are predicted to be at
least second order in $\Omega$ \citep{R89}. The theoretical
predictions are tested by five sets of simulations where $\theta =
45\degr$ is fixed and the rotation rate is varied such that the
Coriolis number covers the range $0.01\ldots0.47$ (see \Table{tab:runs_aniso}).

\begin{figure*}[t!]
  \includegraphics[width=\textwidth]{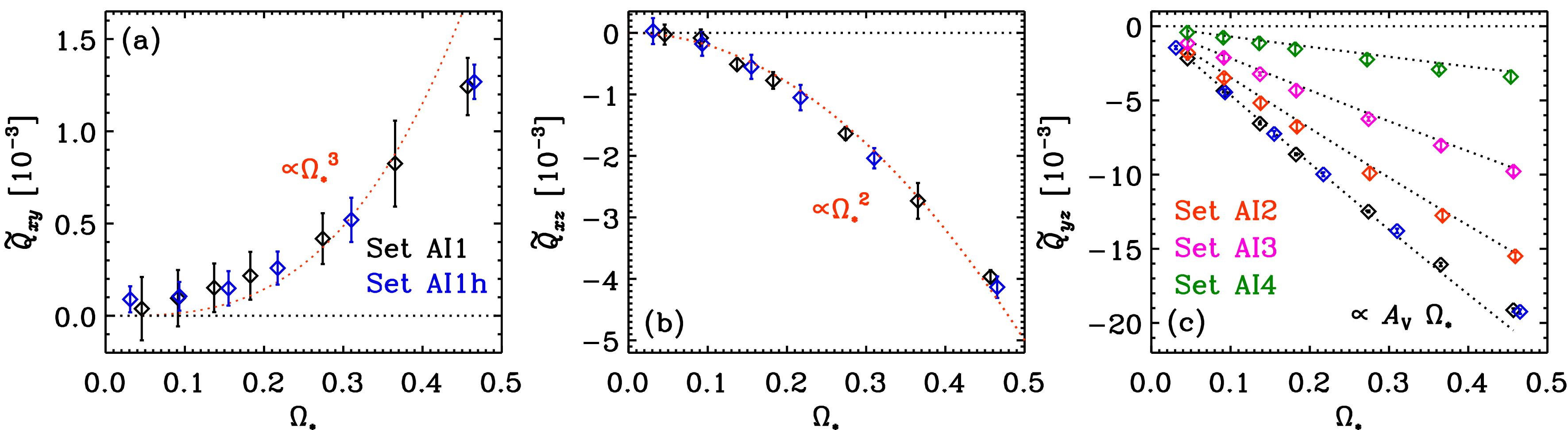}
\caption{Normalized Reynolds stress components $\qxy$, $\qxz$, and
  $\qyz$, panels {\bf a}, {\bf b}, and {\bf c}, respectively, from
  $\theta=45\degr$ as functions of $\Omega_\star$. In panels {\bf a}) and {\bf
    b}) data from Sets~AI1 and AI1h is shown, whereas in {\bf c})
  additional data from Sets~AI2--4 is included. The dotted lines in
  panel {\bf c}) are proportional to $A_{\rm V} \Omega_\star$.}
\label{fig:pstress_Co}
\end{figure*}

\Figu{fig:pstress_Co} shows that the horizontal stress $\qxy$ is
positive and the meridional stress $\qxz$ is negative, indicating
$H>0$ and $M<0$, respectively. The numerical results are consistent
with the analytic results of \cite{R89}, that is $\qxy \propto
\Omega_\star^3$ and $\qxz \propto \Omega_\star^2$. However, due to the
relatively large error of $\tQij{xy}$, it is also compatible with an
$\Omega_\star^2$-dependence. The magnitude of the horizontal stress in
this parameter regime is $\mathcal{O}(10^{-4} \urms^2)$ and the error
estimates are substantial, although the simulations with the smallest
Taylor numbers were integrated for several tens of thousands eddy
turnover times. This is due to the fact that in the slow rotation
limit the horizontal $\Lambda$ effect is proportional to $A_{\rm H}$
\citep{R80}, which is rotation-induced and which remains small in
comparison to $A_{\rm V}$ in this regime (see
\Table{tab:runs_aniso}). It is noteworthy that the meridional stress
$\qxz$ has a clearly greater magnitude in comparison to $\qxy$. This
is due to the stronger $\Omega_\star$ dependence of the latter
\citep{R89}. The vertical stress $\qyz$ is significantly stronger
than the other two components and shows a linear dependence on $\Co$
in accordance with the theoretical prediction (see
\Figu{fig:pstress_Co}c).

\subsection{Dependence of $\Lambda_{\rm V}$ on turbulence anisotropy}

The $\Lambda$ effect depends not only on rotation, but also on the
properties of turbulence \citep{R89}. More specifically, analytic
theories predict that the anisotropy of turbulence plays a crucial
role \citep{R80,R89}. Here the rotation-induced Reynolds stress is
estimated from the Navier-Stokes equations using a minimal $\tau$
approach \citep[for a more complete derivation, see][]{KB08}. The time
derivative of the Reynolds stress is given by
\begin{eqnarray}
\dot{Q}_{ij}^{(\Omega)} = \mean{\dot{u}_i^{(\Omega)}u_j} + \mean{u_i\dot{u}_j^{(\Omega)}}, \label{equ:dotqij}
\end{eqnarray}
where
\begin{eqnarray}
\dot{u}_i^{(\Omega)} = -2 \epsilon_{imn} \Omega_m u_n + N_i, \label{equ:ui}
\end{eqnarray}
where $N_i$ encompasses viscous and non-linear terms. Using \Eq{equ:ui}
in \Eq{equ:dotqij} yields
\begin{eqnarray}
\dot{Q}_{ij}^{(\Omega)} = -2 \epsilon_{jkl} \Omega_k Q_{il} -2 \epsilon_{ikl} \Omega_k Q_{jl} + \mathcal{T}_{ij},
\end{eqnarray}
where $\mathcal{T}_{ij}$ contains triple and higher-order
correlations. Assuming a stationary state with $\dot{Q}_{ij}= 0$, and
approximating the triple correlations as
$\mathcal{T}_{ij}=-Q_{ij}/\tau$ in accordance with the $\tau$
approximation \citep[e.g.][]{2002PhRvL..89z5007B,2003PhFl...15L..73B},
where $\tau$ is a relaxation time, gives
\begin{eqnarray}
Q_{ij}^{(\Omega)} = -2 \tau \epsilon_{jkl} \Omega_k Q_{il} -2 \tau \epsilon_{ikl} \Omega_k Q_{jl}.
\end{eqnarray}
Numerical simulations of turbulent passive scalar and magnetic field
transport have yielded support for the validity of the $\tau$
approximation
\citep[e.g.][]{2004PhFl...16.1020B,2005A&A...439..835B,2012PhyS...86a8406S}.
However, one must bear in mind that this is a rather simplistic
approach to the turbulence closure problem and that the results should
not be considered exact. Associating this with the $\Lambda$ effect,
that is assuming $Q_{ij}^{(\Omega)}=Q_{ij}^{(\Lambda)}$, the vertical
$(yz)$ component of the stress is
\begin{eqnarray}
Q_{yz}^{(\Lambda)} = 2 \tau \Omega_x (\qzz-\qyy) - 2\tau \Omega_z \qxz. \label{equ:qyzL}
\end{eqnarray}
The last term on the right-hand side can be omitted in the slow
rotation limit where $|\qxz| \ll |\qzz-\qyy|$. Furthermore, the
relaxation time can be related to the turnover time $\ell/\urms$ via
$\tau=\St \ell/\urms$, where $\St$ is the Strouhal number. Inserting
this into \Eq{equ:qyzL} and dividing by $\urms^2$ yields
\begin{eqnarray}
\tilde{Q}_{yz}^{(\Lambda)} \approx -\St \Omega_\star A_{\rm V} \sin \theta. \label{equ:qyzAv}
\end{eqnarray}
Four sets of simulations (Sets~AI1--4) were made where the anisotropy
of the turbulence was systematically reduced in comparison to the
maximum case by reducing the ratio of the forcing amplitudes
$f_1/f_0$. Set~AI1h has otherwise  similar parameters to those of AI1, but
was done with forcing wavenumber $\tilde{\kf}=30$ (see
\Table{tab:runs_aniso}).  The runs in Sets~AI2--4 were not integrated as
long as those in Sets~AI1 and AI1h, leading to poor convergence of the
stress components $\qxy$ and $\qxz$. Thus, these results are not shown
here. The vertical stress $\qyz$ is shown as a function of
$\Omega_\star$ from Sets~AI1--4 in \Fig{fig:pstress_Co}(c). The
numerical results indicate that the stress, and thus the vertical
$\Lambda$ effect, is linearly proportional to the turbulence
anisotropy in the slow rotation regime. Comparison of the stress with
\Eq{equ:qyzAv} shows good agreement for all sets of runs with
$\St=0.13$.

\subsection{Reynolds stress and $\Lambda$ effect as functions of rotation}

In a stellar convective envelope the rotational influence on the flow
can vary by several orders of magnitude as a function of radius due to
the strong density stratification. This can be seen from the Coriolis
number for the Sun
\begin{equation}
\Co_{\odot} = 2\,\Omega_\odot \tau = \frac{2\,\Omega_\odot \Hp}{u_{\rm conv}},
\end{equation}
where $\Omega_\odot=2.7\cdot10^{-6}$~s$^{-1}$ is the mean solar
rotation rate, $\tau=\Hp/u_{\rm conv}$ is an estimate convective
turnover time, $\Hp=-(\pd \ln p/\pd r)^{-1}$ is the pressure scale
height, and $u_{\rm conv}$ is the convective rms-velocity. Mixing
length models of the solar convection zone \citep[e.g.][]{Stix02}
yield values of $u_{\rm conv}$ and $\Hp$ such that $\Co_\odot$ ranges
from $10^{-3}$ in the photosphere to roughly unity at
$r=0.95R_\odot$, while reaching values of more than ten at the base of
the CZ \citep[e.g.][]{KKST05}. At least this range in Coriolis numbers
needs to be probed for the results to be usable in mean-field models
of solar and stellar differential rotation.

\begin{table}
\centering
\caption[]{Summary of runs with varying $\Omega_\star$.}
  \label{tab:runs_rot}
       \vspace{-0.5cm}
      $$
          \begin{array}{p{0.05\linewidth}ccccccccccc}
          \hline
          \hline
          \noalign{\smallskip}
          Set & \Ta\ [10^7] & A_{\rm V} & A_{\rm H}\ [10^{-3}] & \Omega_\star \\
          \hline
          A  &  0.06 & -0.53  & \ \ \ \ 0 \ldots 1 & 0.09 \\ %
          B  &  0.25 & -0.52  & \ \ \ \ 0 \ldots 3 & 0.18 \\ %
          C  &  0.56 & -0.52  & \ \ \ \ 0 \ldots 5 & 0.27 \\ %
          D  &  1.6  & -0.51  & \ \ \ \ \ \ 0 \ldots 11 & 0.46 \\ %
          E  &  6.2  & -0.47 \ldots -0.49 & \ \ \ -0 \ldots 28 & 0.91 \\ %
          F  &  25   & -0.39 \ldots -0.43 & \ \ \ -3 \ldots 56 & 1.8 \\ %
          G  &  56   & -0.34 \ldots -0.38 & \ \ -11 \ldots 66 & 2.7 \\ %
          H  &  156  & -0.24 \ldots -0.32 & \ \ -26 \ldots 66 & 4.6 \\ %
         (I  &  623  & -0.25 \ldots+0.61  & -330 \ldots 50 & 5.2 \ldots 9.0) \\ %
          \hline
          AA &  0.39 & -0.60  & \ \ \ \ \ \ 0 \ldots 10 & 0.57 \\ %
          BB &  1.6  & -0.56  & \ \ \ -0 \ldots 25 & 1.1 \\ %
          CC &  6.2  & -0.48 \ldots -0.50 & \ \ \ -0 \ldots 58 & 2.3 \\ %
          DD &  39   & -0.29 \ldots -0.38 & \ \ -25 \ldots 80 & 5.7 \\ %
          EE &  156  & -0.21 \ldots -0.29 & \ \ -29 \ldots 59 & 11 \\ %
          FF &  623  & -0.18 \ldots -0.25 & \ \ -15 \ldots 36 & 23 \\ %
          GG &  2490 & -0.14 \ldots -0.24 & \ \ \ -0 \ldots 37 & 45 \\ %
          \hline
          \end{array}
          $$ \tablefoot{ All runs have $f_0=10^{-6}$,
            $f_1/f_0=4\cdot10^4$, $\tilde{k}_{\rm f}=10$, and grid
            resolution $144^3$. The fluid Reynolds number is $14$ in
            Sets~A--H and $5.5$ in Sets~AA--GG. In Set~I the Reynolds
            number varies in the range $14\lesssim \Rey \lesssim
            24$. Set~I is listed in brackets for completeness but
            the data is not used in the analysis due to the
           occurance of large-scale flows.}
\end{table}

\subsubsection{Latitudinal dependence of the Reynolds stress}
\label{subsec:latdep}

\begin{figure}
  \includegraphics[width=\columnwidth]{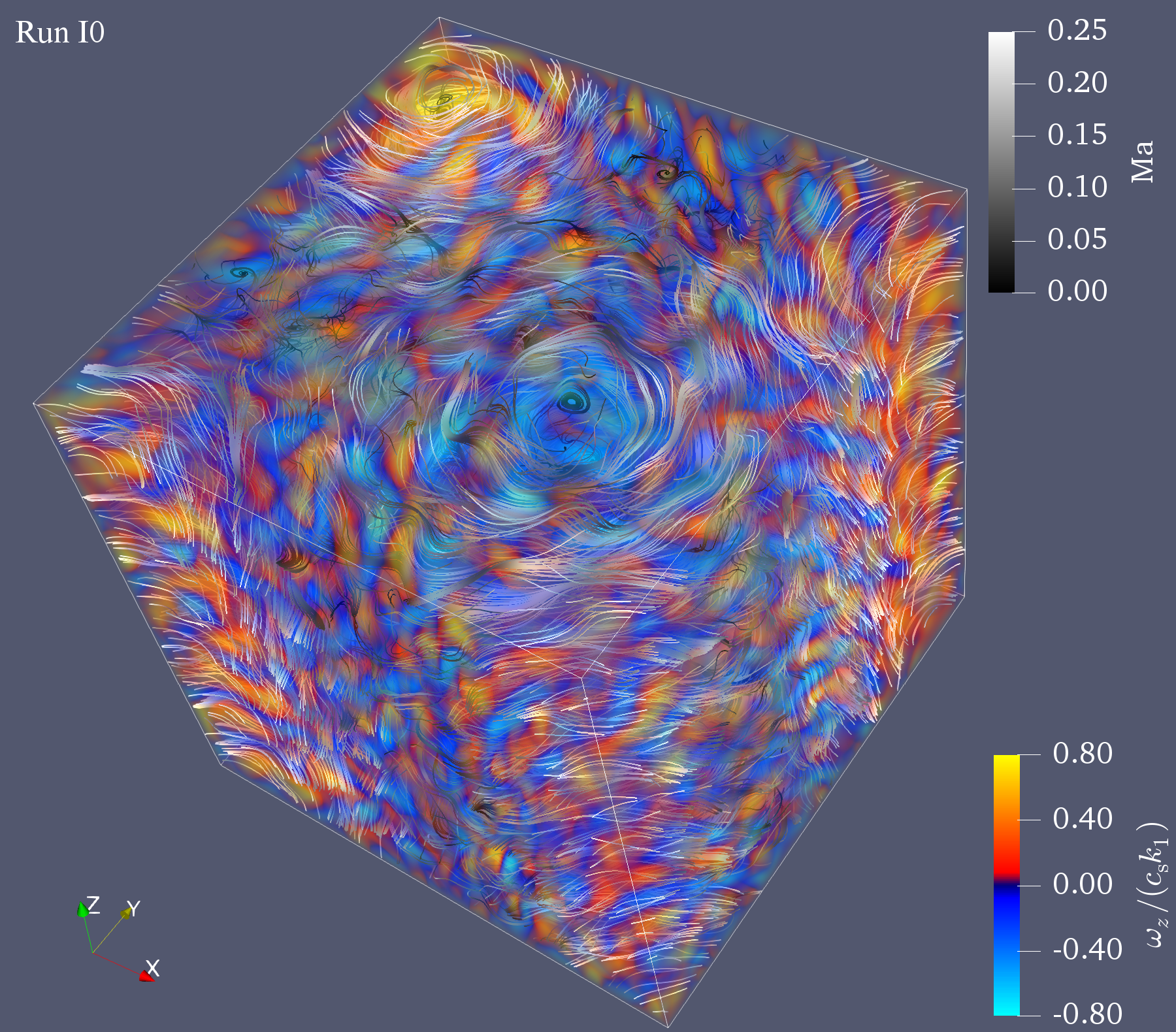}
  \caption{Vertical component of vorticity $\omega_z = (\bm\nabla
    \times {\bm U})_z$ in units of $\cs k_1$ (colour contours) and
    flow vectors (black and white streamlines) in terms of the local
    Mach number $\Ma=|{\bm u}|/c_s$ from Run~I0.}
\label{fig:H0_img_0403}
\end{figure}

The runs probing the rotation dependence are listed in
\Table{tab:runs_rot}. The Reynolds numbers in these runs are
relatively modest\footnote{The Reynolds number based on $\kf$ is
  a factor of $2\pi$  smaller than the usually adopted definition with a length scale
  $\ell=2\pi/\kf$.}
($\Rey=5.5\ldots24$). However, the Reynolds number is the ratio of the
viscous ($\tau_\nu$) to turbulent turnover ($\tau_{\rm u}$) times,
that is $\Rey=(\urms \kf)(1/\nu \kf^2)=\tau_\nu/\tau_{\rm u}$. Thus,
the viscous timescale is always significantly longer than the
characteristic flow timescale in the current simulations.  The values
of $\Omega_\star$ in Sets~A--I range from $0.09$ to $9,$ which roughly
corresponds to the range expected in the solar convection
zone. However, for the fiducial value of the Reynolds number
($\Rey=14$), the flow develops a large-scale vortex at $\theta=0$ when
the Taylor number is increased from $1.56\cdot10^9$ to $6.23\cdot10^9$
corresponding to $4.5 \lesssim \Omega_\star \lesssim 9.0$ (see
\Fig{fig:H0_img_0403}). Similar vortices have been obtained in the rapid
rotation regime in other settings where the turbulence is driven
either by compressible or Boussinesq convection
\citep[e.g.][]{Chan03,Chan07,2011ApJ...742...34K,GHJ14,2014PhRvL.112n4501R}
or isotropic forcing similar to the current study
\citep[e.g.][]{2016PhRvX...6d1036B}. These structures dominate the
flow in the statistically saturated state which explains the extreme
values of $A_{\rm V,H}$, $\Omega_\star$, and $\Rey$ in Set~I (see
\Table{tab:runs_rot}). In cases where the rotation vector is inclined
with the direction of anisotropy, mesoscale flow structures with modes
such as $(k_x,k_y,0)=(3,0,0)$ for $U_y$ appear. Similar structures also
dominate the Reynolds stress and overwhelm the turbulent
contributions. Thus, Set~I is disregarded from further analysis and
higher rotation rates were not explored with $\Rey=14$. Instead, a
series of runs were made with $\Rey=5.5$ (Sets~AA--GG in \Table{tab:runs_rot}),
roughly overlapping with Sets~D--H,  to study the rapid
rotation regime. Vortices do not appear in these runs even at
substantially higher rotation rates (up to $\Omega_\star\approx
45$). This is consistent with the finding of
\cite{2011ApJ...742...34K} that a critical Reynolds number has to be
exceeded for the vortices to form. The study of large-scale vorticity
generation and its effects on angular momentum transport will be
presented elsewhere.

\begin{figure}
  \includegraphics[width=\columnwidth]{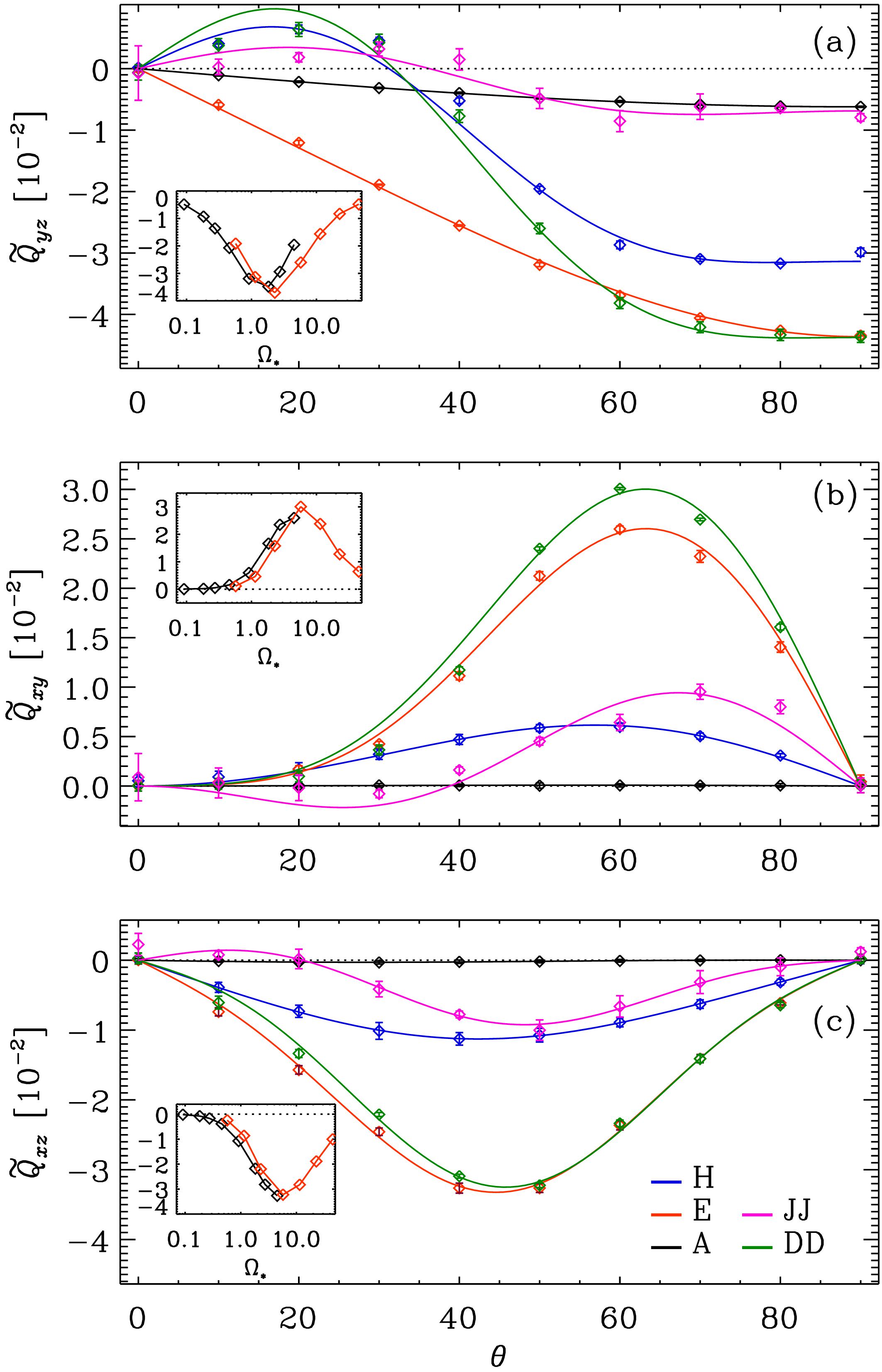}
\caption{Off-diagonal stresses from a representative set of runs.
    Diamonds: normalized Reynolds stress components {\bf a})
    $\tQij{yz}$, {\bf b}) $\tQij{xy}$, and {\bf c}) $\tQij{xz}$ as
    functions of $\theta$ from Sets~A (black), E (red), H (blue), AA
    (green), and GG (magenta). Curves: fits with the $\Lambda$
    coefficients presented in \Seca{sec:Lampara}. The insets show the
    corresponding stresses from the respective latitudinal maxima {\bf
      a}) $\theta=90\degr$, {\bf b}) $\theta=60\degr$, and {\bf c})
    $\theta=50\degr$ as functions of $\Omega_\star$, where data from
    Sets~A--H (black) and AA--GG (red) are indicated.}
\label{fig:pqij_Co}
\end{figure}

Figure~\ref{fig:pqij_Co} shows the off-diagonal stresses $\qyz$,
$\qxy$, and $\qxz$ from a representative selection of
runs. In the case of slowest rotation (Set~A, $\Omega_\star=0.09$),
the horizontal stress $\qxy$ is not statistically significant, whereas
the meridional stress $\qxz$ is barely so (see also
\Figa{fig:pstress}). The vertical stress $\qyz$ is well-defined due to
the strong anisotropy of the turbulence already at the lowest rotation
rate considered here. At more rapid rotations the horizontal
(meridional) stress acquires consistently positive (negative) values
at all latitudes (Set~E with $\Omega_\star = 0.91$;
Fig.~\ref{fig:pqij_Co}b). The latitude at which the horizontal and
meridional stresses peak shifts toward the equator as a function of
$\Omega_\star$. In the regime of rapid rotation, the maximum of
$\qxy$ continues to
move to lower latitudes, but this trend is not as extreme as in local
simulations of rapidly rotating convection
\citep[e.g.][]{KKT04,2005AN....326..315R,2005AN....326..223H}. The
reason is that in convection simulations large-scale flow structures,
also known as banana cells \citep{1970ApJ...159..629B}, develop near
the equator and enhance the horizontal stress \citep{KMGBC11}. However, such
flow structures are absent in the current simplified models.
In Sets~FF and GG, with $\Omega_\star=23$ and $45$, $\qxy$ shows
indications of a sign change at high latitudes, which was not found by
\cite{KB08}\footnote{The Coriolis number in \cite{KB08} differs from
  the current definition by a factor of $2\pi$, that is
  $\Omega_\star=2\pi\Co$.}. The meridional stress $\qxz$ reaches a
maximum around $\theta=45\degr$ at intermediate rotation
($\Omega_\star=0.5\ldots10$) and no clear sign change is observed even
at higher $\Omega_\star$. Also, this differs from the results of
\cite{KB08} where a sign change occurred near the equator for
$\Omega_\star\approx 34$ in the current units. In the most rapidly
rotating runs (Set~GG), some indication of a sign change at high
latitudes is present. Formally, the differences of the current
Sets~A--I to the runs of \cite{KB08} are minor: the grid resolution is
roughly twice as high in the current simulations, and the forcing is
applied at $\tilde{k}_{\rm f}=10$ instead of $\tilde{k}_{\rm
  f}=5$. Finally, the Mach number is roughly  a factor of
three smaller in the current simulations. It is unclear which of the
differences is causing the results to diverge at rapid rotation. The
results at slow and intermediate rotation ($0.09\lesssim \Omega_\star
\lesssim 20$) are, however, in good agreement with those of
\cite{KB08}.

The vertical stress $\qyz$ is clearly the dominant component in the
current models. In the slow rotation regime it shows a stable
configuration such that the values are consistently negative, and the
latitude dependence experiences only minor changes until around
$\Omega_\star \approx 0.9$ (see \Figa{fig:pqij_Co}). This is consistent
with theory of the vertical $\Lambda$ effect at slow rotation
\citep{R89}. In the rapid rotation regime a sign change occurs at high
latitudes and the magnitude is drastically reduced near the equator in
accordance with the results of \cite{KB08} (see the inset in
\Figa{fig:pqij_Co}a). A low latitude sign change of the vertical
stress is not observed in contrast to local and global convection
simulations where also the vertical turbulence anisotropy changes sign
\citep[e.g.][]{KKT04,KKB14}. The magnitudes of both $\qxy$ and $\qxz$
increase until $\Omega_\star\approx 5$ with the latter being somewhat
larger at all rotation rates (see the insets in
Figs.~\ref{fig:pqij_Co}b and c). The decrease in the stress at rapid
rotation is associated with rotational quenching of the $\Lambda$
effect \citep{KR93}, which is likely due to the reduced anisotropy in
that regime (see \Table{tab:runs_rot}). Differences with convection
simulations may be explained by the missing contributions from the
heat flux in the present models \citep{2006PhRvE..73d6303K}.

\begin{figure}
  \includegraphics[width=\columnwidth]{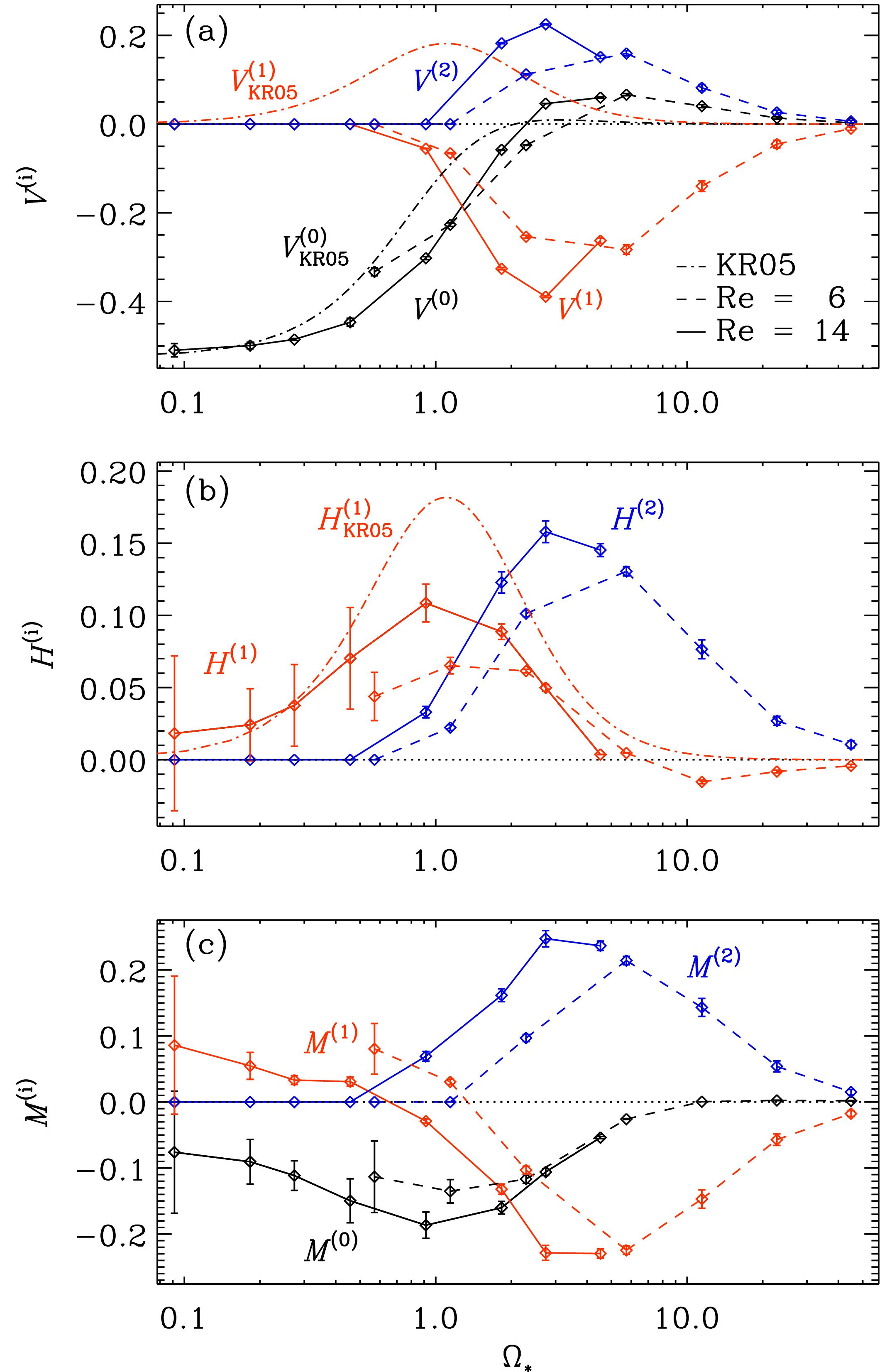}
\caption{Coefficients $V^{(i)}$, $H^{(i)}$, and $M^{(i)}$ as functions
  of $\Omega_\star$ from Sets~A--H (solid lines) and AA--GG
  (dashed). The dash-dotted lines in {\bf a}) and {\bf b}) correspond
  to analytic results \Eqsa{equ:V0SOCA}{equ:H1SOCA} as indicated by
  the legends. Here $a=1.75$ was used to match $V^{(0)}=V^{(0)}_{\rm
    KR05}$.}
\label{fig:pLambda}
\end{figure}

\subsubsection{Parameterization in terms of the $\Lambda$ effect}
\label{sec:Lampara}

The procedure outlined in \Seca{sec:strategy} was used to extract the
coefficients pertaining to the $\Lambda$
effect. Figure~\ref{fig:pLambda} shows the results obtained from
Sets~A--H and AA--GG. The fundamental mode of the $\Lambda$ effect is
recovered in the  slow rotation limit
($\Omega_\star\lesssim0.5$), manifested by $V^{(0)}$ tending to a
constant value of $\approx-0.5$ with $V^{(1)}$ and $V^{(2)}$ being
negligible. The vertical stress is well described by $V^{(0)}$ until
$\Omega_\star\approx 1$, beyond which the higher-order components are
needed. In the fits obtained, $V^{(1)}$ and $V^{(2)}$ have different
signs with the former having roughly twice the absolute magnitude of
the latter. Reasonable agreement is found between the overlapping sets
of runs with different values of $\Rey$ (Sets~D--H and AA--DD) in the
regime of intermediate rotation ($0.5\lesssim \Omega_\star \lesssim
5$). In the rapid rotation regime all of the $V^{(i)}$ coefficients
are quenched and tend to almost zero at the highest value of
$\Omega_\star$. The value of $V^{(0)}$ found in the slow rotation
regime is almost exactly half  that indicated by helioseismology
\citep{2014A&A...570L..12B}. The current results, however, depend
crucially on the adopted value of $\nut$, see
\Equsa{equ:Vsim}{equ:fancyR}. By using $\nut={2 \over 15} \urms
\ell$ \citep[e.g.][]{R89}, the observational result $V^{(0)}\approx-1$
would be recovered. This highlights the arbitrariness of the choice of
$\nut$ and the need for methods to estimate it independently.

\begin{figure}
  \includegraphics[width=\columnwidth]{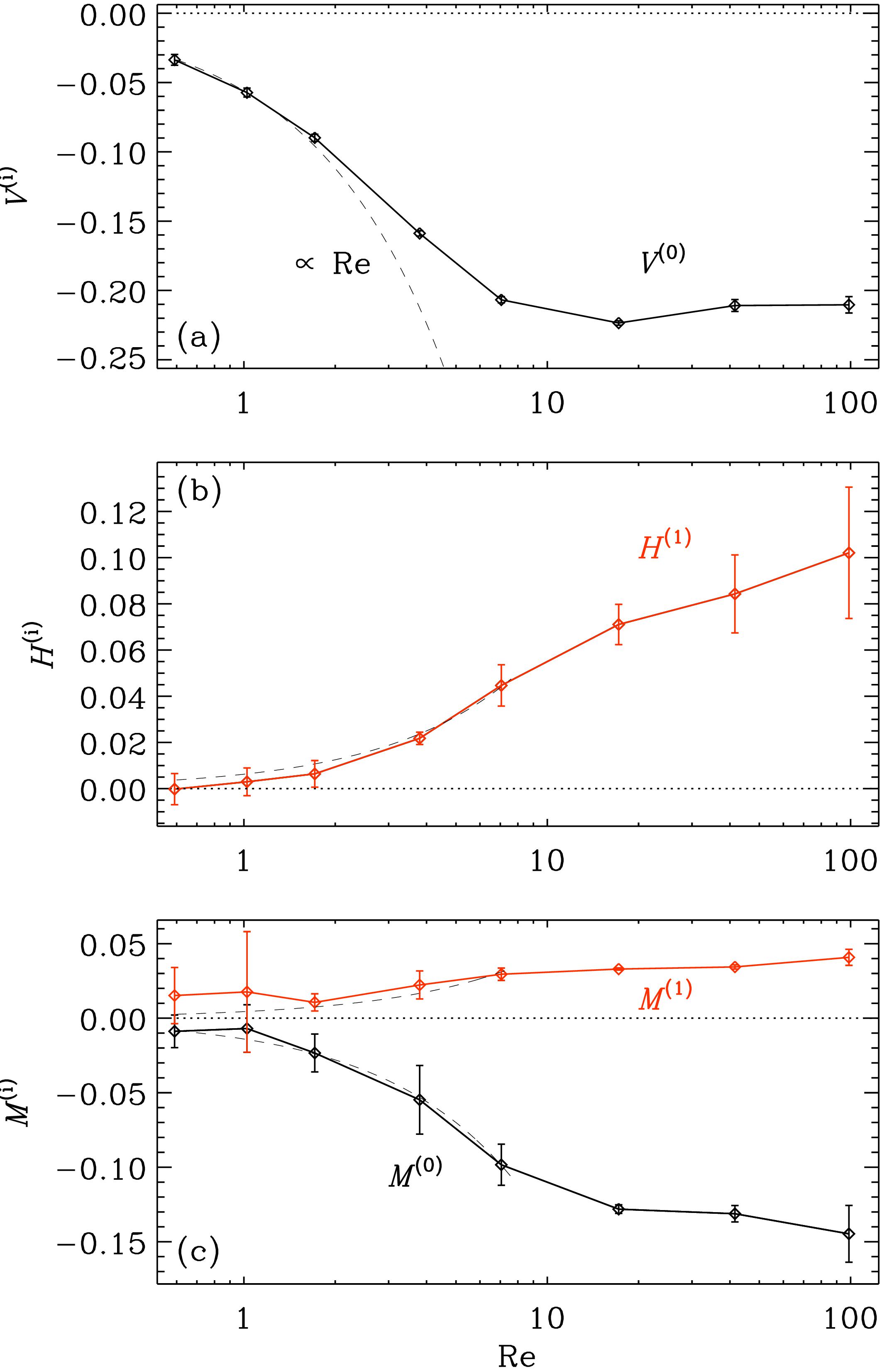}
\caption{Coefficients pertaining to the $\Lambda$ effect as functions
  of $\Rey$: $V^{(0)}$ ({\bf a}), $H^{(1)}$ ({\bf b}), and $M^{(0,1)}$
({\bf c}) for Sets~RE1--RE8 with $\Omega^\star=1.0$ and $A_{\rm
    V}\approx-0.3$. The dashed lines are proportional to $\Rey$.}
\label{fig:pLambda_re}
\end{figure}

The coefficients $H^{(1)}$ and $H^{(2)}$, corresponding to the
horizontal $\Lambda$ effect, are always positive in Sets~A--H.  The
results from Sets~AA--CC agree qualitatively with the higher-$\Rey$
runs although the values are generally lower. This is due to the fact
that the coefficients are not yet in an asymptotic regime with
respect to the Reynolds number (see \Seca{sec:reydep} and
\Fig{fig:pLambda_re}). In the lower-$\Rey$ runs $H^{(1)}$ turns
negative around $\Omega_\star\approx 11$. For $\Omega_\star \gtrsim 5$
a strong rotational quenching is observed and the $H^{(i)}$
coefficients also tend to very small values at the highest rotation
rates corresponding to $\Omega_\star=23$ and $45$.

The hitherto poorly studied meridional $\Lambda$ coefficients are
shown in \Fig{fig:pLambda}c). The simple $\cos\theta \sin\theta$
dependence of the stress is well described by $M^{(0)}$ alone at slow
rotation ($\Omega_\star \lesssim 0.5$). For $\Omega_\star \gtrsim
0.5$, this behaviour gives way to a stronger concentration at
mid-latitudes. This is manifested by a diminishing $M^{(0)}$ with
increasing $M^{(1)}$ and $M^{(2)}$ with almost equal absolute
magnitudes but opposite signs for $\Omega_\star \gtrsim 1$. As with
the vertical and horizontal $\Lambda$ effects, a strong rotational
quenching is observed for $\Omega_\star \gtrsim 5$.  The
correspondence between the lower and higher Reynolds number runs is
again reasonably good.

\subsubsection{Comparison to analytic results}

In the studies of \cite{2004ARep...48..153K} and \cite{KR05} a
distinction is made between the contributions from density
stratification and anisotropy of turbulence to the $\Lambda$
effect. The latter is dominant in the slowly rotating regime which
corresponds to the upper layers of the solar convection zone and the
current simulations. The analytic model of \cite{KR05} predicts the
following functional forms for $V^{(0)}$ and $H^{(1)}$ (with
$V^{(1)}=H^{(1)}$) in the case where the $\Lambda$ effect is solely
due to turbulence anisotropy
\begin{eqnarray}
V^{(0)}_{\rm KR05} &=& a \left( \frac{\ell_{\rm corr}}{H_\rho} \right)^2 I_0(\Cost) \label{equ:V0SOCA} ,\\
H^{(1)}_{\rm KR05} &=& a \left( \frac{\ell_{\rm corr}}{H_\rho} \right)^2 I_1(\Cost) \label{equ:H1SOCA}
,\end{eqnarray}
where $a$ is an anisotropy parameter (see below), $\ell_{\rm corr}$ is
the correlation length of turbulence, and $H_\rho=-(\pd \ln \rho/\pd
r)^{-1}$ is the density scale height. The correlation length
$\ell_{\rm corr}$ is taken to equal the mixing length by
\cite{2004ARep...48..153K} and \cite{KR05}, that is $\ell_{\rm corr} =
\alpha_{\rm MLT} H_p$, where $\alpha_{\rm MLT}=1.7$ is the mixing
length parameter and $H_p$ is the pressure scale height. In the
present case $\ell_{\rm corr}$ is taken to correspond to the forcing
scale of turbulence $\ell=2\pi/\kf$, but no scale corresponding to
$H_\rho$ can be identified due to the homogeneity of the system under
consideration. Furthermore, $H_\rho=L_{\rm d}$ is assumed for
simplicity. The quenching functions $I_0$ and $I_1$ are given by
\begin{eqnarray}
I_0 &=& \frac{1}{4\Cost^4} \left( -19 - \frac{5}{1+\Cost^2} + \frac{3\Cost^2+24}{\Cost} \arctan \Cost \right)\!, \\
I_1 &=& \frac{3}{4\Cost^4} \left( -15 - \frac{5\Cost^2}{1+\Cost^2} + \frac{3\Cost^2+15}{\Cost} \arctan \Cost \right)\!.
\end{eqnarray}

The analytic results $V^{(0)}_{\rm KR05}$ and $H^{(1)}_{\rm KR05}$ are
compared with the numerically obtained coefficients $V^{(i)}$ and
$H^{(1)}$ in \Figa{fig:pLambda}. The analytic and numerical results
are in rough qualitative agreement for slow rotation ($\Co\lesssim1$),
but several differences are immediately apparent. First, the
numerical data is at odds with the analytic result indicating that
$V^{(1)}=H^{(1)}$. An obvious candidate for the discrepancy is
that
the numerical values are obtained by fitting where $V^{(1)}$ and
$H^{(1)}$ are considered independent. The second major difference is
that the numerical data for sufficiently rapid rotation ($\Omega_\star
\gtrsim 1$) is incompatible with expressions of $V$ and $H$ which
consider only terms proportional to $\sin^2\theta$.

\begin{table}
\centering
\caption[]{Summary of runs where $\Rey$ was varied.}
  \label{tab:runs_rey}
       \vspace{-0.5cm}
      $$
          \begin{array}{p{0.05\linewidth}cccccccccccc}
          \hline
          \hline
          \noalign{\smallskip}
          Set & f_0\ [10^{-2}] & f_1/f_0 & \Ta\ [10^7] & A_{\rm H}\ [10^{-3}] & \Rey \\
          \hline
          RE1  &  5.5 & 3.3 & 0.014 & -0 \ldots 3 & 0.6 \\ %
          RE2  &  5.5 & 3.5 & 0.04  & -0 \ldots 5 & 1.0 \\ %
          RE3  &  5.5 & 3.6 & 0.12  & -0 \ldots 4 & 1.7 \\ %
          RE4  &  5.5 & 4.2 & 0.57  & -0 \ldots 8 & 3.8 \\ %
          RE5  &  5.5 & 4.9 & 2.0   &  \ \ \ \ \ 0 \ldots 13 & 7.1 \\ %
          RE6  &  4.8 & 6.5 & 12    &  \ \ \ \ \ 0 \ldots 18 & 17 \\ %
          RE7  &  4.0 & 10  & 69    & \ \ -0 \ldots 19 & 42 \\ %
          RE8  &  10  & 44  & 409   &  \ \ \ \ \ 0 \ldots 23 & 99 \\ %
          \hline
          \end{array}
          $$ \tablefoot{ All runs have $\tilde{k}_{\rm f}=10$, $A_{\rm
              V}=-0.26 \ldots -0.31$, and $\Omega_\star=1.0$. Grid
            resolutions $144^3$ (Sets~RE1--6), $288^3$ (RE7), and
            $576^3$ (RE8) were used.}
\end{table}

Furthermore, in the foregoing analysis the anisotropy parameter $a$
was kept as a free parameter and tuned such that $V^{(0)}_{\rm
  KR05}=V^{(0)}$ in the slow rotation limit. It is also possible to
compute $a$ directly using  Eq.~A14 in Appendix~A of \cite{KR05} by
substituting $u_r^2 = \mean{u_z^2}$ and $u_\theta^2 =u_\phi^2 = u_{\rm
  H}^2\equiv\onehalf(\mean{u_x^2}+\mean{u_y^2})$, and defining
$b\equiv\mean{u_z^2}/\mean{u_H^2}$:
\begin{eqnarray}
  a = \frac{ b \left( 1+\frac{3}{2}\frac{\ell_{\rm corr}^2}{L^2} \right)-1}{\frac{\ell_{\rm corr}^2}{L^2}(\frac{2}{5}+\frac{b}{5})}, \label{equ:a}
\end{eqnarray}
where $L$ is a length scale corresponding to large-scale
inhomogeneity. Assuming $L=L_{\rm d}$ and $\ell_{\rm corr}=\ell$ gives
$\ell_{\rm corr}/L_{\rm d} = \ell/L = k_1/\kf\approx 0.1$, allowing
\Eq{equ:a} to be solved with $b$ as an input from simulations. The
maximum anisotropy used in the bulk of the simulations is $A_{\rm
  V}\approx0.52$, which yields $b\approx2.1$ and $a\approx2.7$. This
is somewhat greater than the values used in the fitting
above. However, this discrepancy is mostly due to the freedom in
choosing the value of $\nut$.

\subsection{Dependence of $\Lambda$ effect on Reynolds number}
\label{sec:reydep}

The simulations in the preceding sections were made at low Reynolds
numbers in comparison to the astrophysically relevant
regime. \Figu{fig:pLambda_re} shows the $\Lambda$ coefficients for a
representative case where the Coriolis number
($\Omega^\star\approx1.0$) and turbulence anisotropy ($A_{\rm
  V}\approx0.3$) were fixed in the range $\Rey\approx0.6\ldots
99$. The Coriolis number was chosen such that the higher-order
coefficients $V^{(1)}$, $V^{(2)}$, $H^{(2)}$, and $M^{(2)}$ did not
appear according to the criteria in \Sec{sec:strategy}. At low Reynolds
numbers $\Rey=0.6\ldots2$ all of the coefficients are proportional to
$\Rey$ as expected from mean-field theory. At low $\Rey$ the error
bars for $H^{(1)}$ and $M^{(0,1)}$ increase because the mean values of
$\qxy$ and $\qxz$ are small. Furthermore, the coefficients level off
beyond $\Rey\gtrsim10$ where they are consistent with constants
although with large error bars in particular for $H^{(1)}$. These
results suggest that the results obtained at $\Rey\approx10$ are
representative of what can be expected in more turbulent cases.

\subsection{Influence of large-scale magnetic fields}

Dynamically significant magnetic fields are ubiquitous in
astrophysical objects where the $\Lambda$ effect is thought to be
important. In particular, stars with convection zones harbour dynamos
that produce magnetic fields on various scales. The effects of
large-scale magnetic fields on the $\Lambda$ effect have been studied
analytically by \cite{KRK94} and \cite{Ki16}. These studies indicate
that large-scale magnetic fields tend to quench the $\Lambda$ effect,
but also that an additional $H^{(0)}$ effect arises in the presence of
a horizontal mean magnetic field. Here the study of \cite{KRK94} is
followed where the quenching formulae for $V^{(0)}$ and $H^{(0)}$ were
calculated as functions of the magnetic field assuming the field to be
horizontal. They found that
\begin{eqnarray}
V^{(0)}_{\rm KPR94} &=& K_1 (\beta) \mathcal{G}, \label{equ:V0KPR94} \\
H^{(0)}_{\rm KPR94} &=& K_2 (\beta) \mathcal{G}, \label{equ:H0KPR94}
\end{eqnarray}
where
\begin{eqnarray}
K_1 &=& \frac{1}{16\beta^4} \left( \frac{\beta^2+1}{\beta}\arctan \beta - 1 - \frac{2\beta^2}{3(1+\beta^2)} \right)\!, \\
K_2 &=& \frac{1}{16\beta^4} \left( -15 - \frac{5\beta^2}{1+\beta^2} + \frac{3\beta^2+15}{\beta} \arctan \beta \right)\!,
\end{eqnarray}
$\beta=B_0/\Beq$, and
\begin{eqnarray}
\mathcal{G} = \tau_{\rm corr}^2 \frac{\pd^2 \overline{\uuu^2}}{\pd r^2}.
\end{eqnarray}
These equations correspond to the case where the anisotropy is due to
the density stratification and where $\mathcal{G}$ is describing
this. These equations indicate that $V^{(0)}$ is monotonically
quenched by magnetic fields, whereas $H^{(0)}$ vanishes as
$\beta\rightarrow0$ and obtains a maximum for $\beta\approx0.94$.
Direct comparison to the analytic study is not possible since the
turbulence intensity is homogeneous in the current simulations. Thus,
$\mathcal{G}$ is treated here as a free parameter.

Here the dependence of the $\Lambda$ effect on large-scale magnetic
fields is studied systematically with controlled numerical experiments,
where either a uniform horizontal (Sets~LSFH1-9) or a vertical
(Sets~LSFV1-9) large-scale imposed magnetic field is present (see
\Table{tab:runs_mag}). The magnetic Reynolds number ($\ReM\approx14$)
is chosen such that it does not exceed the critical value
$\ReM\approx30$ for a small-scale dynamo to be excited \citep{B01}.
The same analysis as above in the hydrodynamic case is performed on
the total turbulent stress,
\begin{eqnarray}
\tij = \qij - \rho^{-1}\mij,
\end{eqnarray}
where $\mij=\mu_0^{-1}\overline{b_i b_j}$ is the Maxwell stress. In
the current fully periodic and homogeneous case no large-scale
magnetic fields, apart from the imposed field $\mean{\bm B}^{(0)}$,
are present. Lundquist numbers ranging from $0.1$ to $50$ are studied
for both field geometries (see \Table{tab:runs_mag}). This range
corresponds to $7\cdot10^{-3} \ldots 3.5$ in terms of the
equipartition strength $\Beq$.

\begin{table}[t!]
\centering
\caption[]{Summary of runs with magnetic fields.}
  \label{tab:runs_mag}
       \vspace{-0.5cm}
      $$
          \begin{array}{p{0.075\linewidth}ccccccccccccc}
          \hline
          \hline
          \noalign{\smallskip}
          Set & f_0\ [10^{-3}] & f_1/f_0 & \Lu & A_{\rm H}\ [10^{-3}] \\
          \hline
          LSFH1 & 10^{-3} & 4\cdot10^4 & 0.1 & \ \ \ 0\ldots 28 \\
          LSFH2 & 10^{-3} & 4\cdot10^4 & 0.2 & -0\ldots 28 \\
          LSFH3 &   0.5   & 80 & 0.5 & -0\ldots 27 \\
          LSFH4 &   1.0   & 39 & 1.0 & -1\ldots 25 \\
          LSFH5 &   2.0   & 19 & 2.0 & -2\ldots 18 \\
          LSFH6 &   3.5   & 11 & 5.0 & \!\! -6\ldots 6 \\
          LSFH7 &   3.5   & 11 & 10  & \!\! -7\ldots 0 \\
          LSFH8 &   3.5   & 10 & 20  & \!\! -7\ldots 5 \\
          LSFH9 &   3.0   & 13 & 50  & -9\ldots 11 \\
          \hline
          LSFV1 & 10^{-3} & 4\cdot10^4 & 0.1 & -0\ldots 28 \\
          LSFV2 & 10^{-3} & 4\cdot10^4 & 0.2 & -0\ldots 28 \\
          LSFV3 & 10^{-3} & 4\cdot10^4 & 0.5 & \ \ \ 0\ldots 28 \\
          LSFV4 &   1.0   &     40     & 1.0 & -0\ldots 24 \\
          LSFV5 &   2.5   &     14     & 2.0 & -0\ldots 19 \\
          LSFV6 &   4.0   &      9     & 5.0 & -0\ldots 11 \\
          LSFV7 &   4.5   &      8     & 10  & \ \  0\ldots 9  \\
          LSFV8 &   4.5   &      8     & 20  & \ \ \ 0\ldots 11 \\
          LSFV9 &   3.5   &     10     & 50  & \ \ \ 0\ldots 21 \\
          \hline
          \end{array}
          $$ \tablefoot{ All runs have $\Ta=6.2\cdot10^7$,
            $\Omega_\star=0.90\ldots0.94$, $A_{\rm
              V}=-0.46\ldots-0.50$, $\Rey=\Rm=13\ldots14$, and grid
            resolution $144^3$.}
\end{table}

\begin{figure}
  \includegraphics[width=\columnwidth]{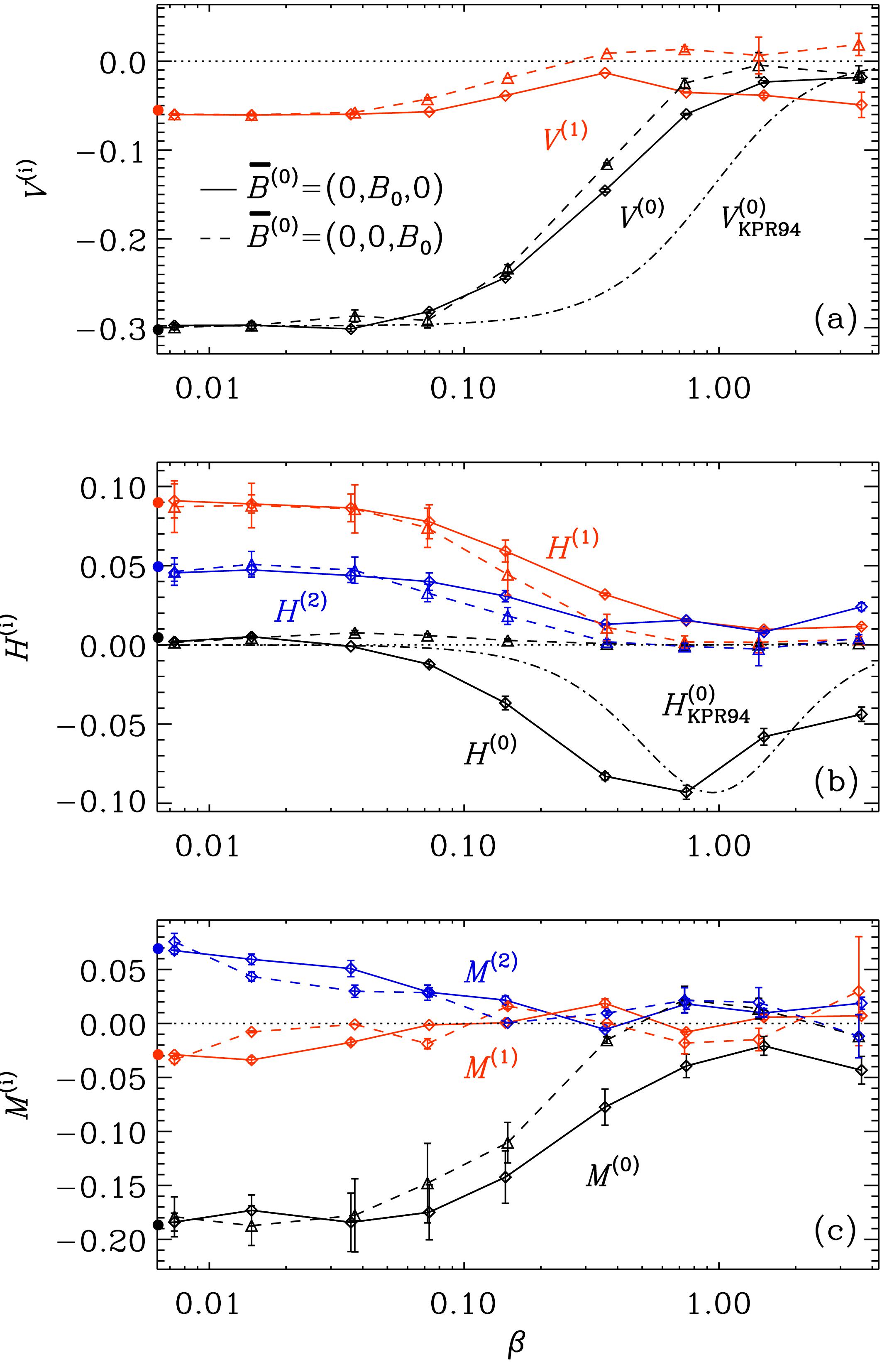}
\caption{Coefficients $V^{(i)}$ ({\bf a}), $H^{(i)}$ ({\bf b}), and
  $M^{(i)}$ ({\bf c}) as functions of $\beta$. Solid (dashed) lines
  correspond to runs with an imposed horizontal (vertical) field. The
  dotted horizontal line denotes the zero level. The open circles on
  the left axis indicate the hydrodynamic values from Set~E. The
  dash-dotted lines show analytic results according to {\bf a})
  \Eq{equ:V0KPR94} (top panel) and {\bf b}) \Eq{equ:H0KPR94}
  (middle).}
\label{fig:pLambda_lsf}
\end{figure}

\begin{figure}
  \includegraphics[width=\columnwidth]{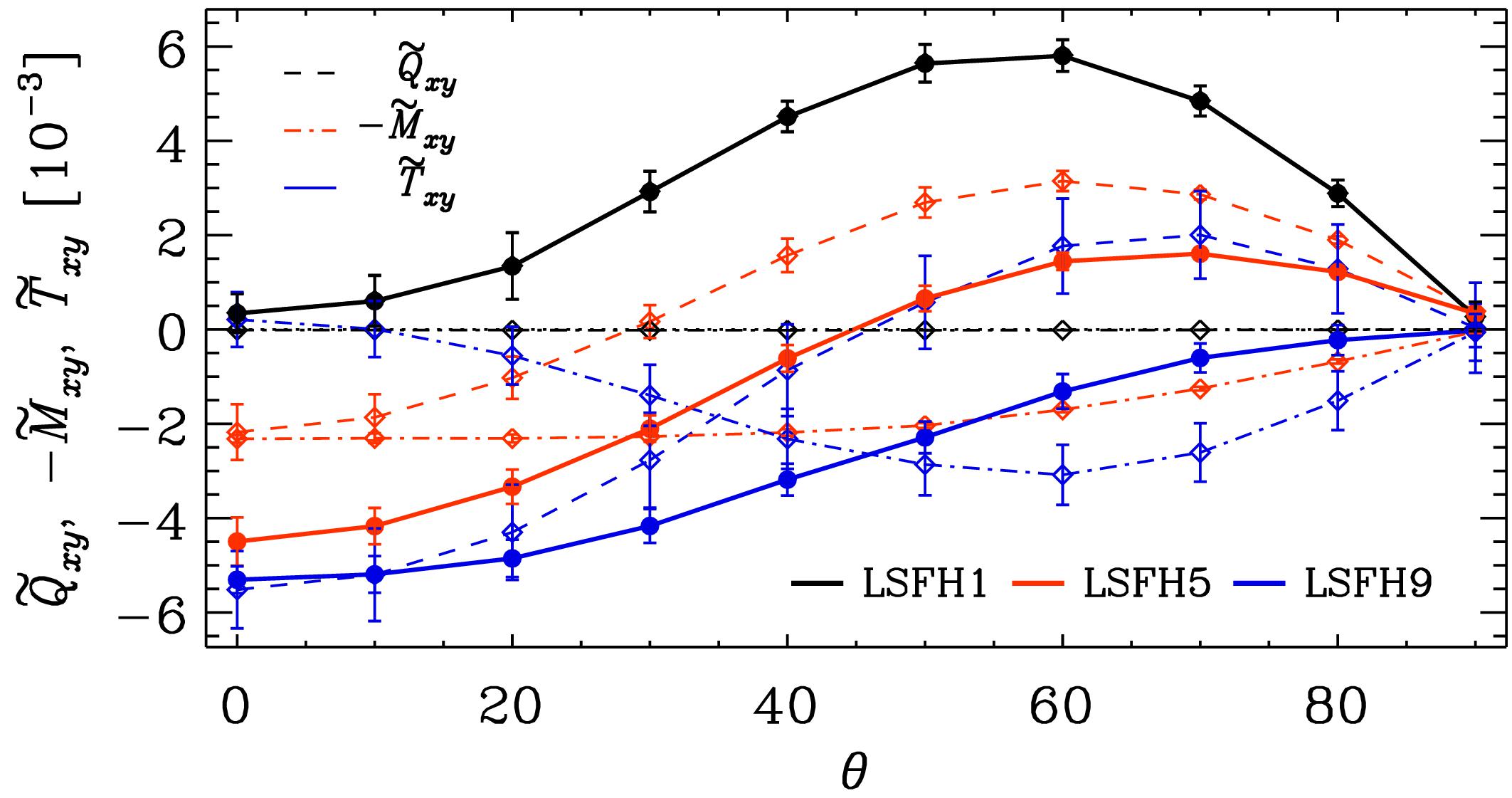}
\caption{Normalized horizontal ($xy$), Reynolds (dashed), Maxwell
  (dash-dotted), and total (thick solid) stress from Sets~LSFH1
  (black), LSFH5 (red), and LSFH9 (blue) as functions of $\theta$.}
\label{fig:pqxy_lsf}
\end{figure}

The $\Lambda$ coefficients obtained from the numerical models are
shown in \Figa{fig:pLambda_lsf}. The effect of the large-scale
magnetic field begins to be noticeable for $\Lu=1$ corresponding to
roughly $0.1\Beq$, although $M^{(1)}$ and $M^{(2)}$ are affected
already by weaker fields. For stronger fields the magnitude of all
coefficients, except $H^{(0)}$ in the runs with a horizontal field,
start to decrease. The existence of a non-zero $H^{(0)}$ was
predicted analytically by \cite{KRK94} and the current simulations
confirm this finding numerically for the first time. This part of the
$\Lambda$ effect arises not only due to a negative contribution from
the Maxwell stress, but also from a gradual sign change of the Reynolds
stress starting from the poles (see \Figa{fig:pqxy_lsf}). In the case
of the strongest imposed field (Set~LSFH9 with
$\beta\approx3.5$) the horizontal stress is negative at all latitudes
apart from the equator. In reality, the stress must vanish at
the poles because $\mean{B}_\phi$ in spherical polar coordinates
(corresponding to $\mean{B}_y$ in the current coordinates) also
vanishes. The analytic results, where $\mathcal{G}$ is tuned to match
the numerical results of $V^{(0)}$ at $\beta=0$ and the maximum
amplitude of $H^{(0)}$, are shown alongside the numerical results in
\Figa{fig:pLambda_lsf}.

\begin{figure}
  \includegraphics[width=\columnwidth]{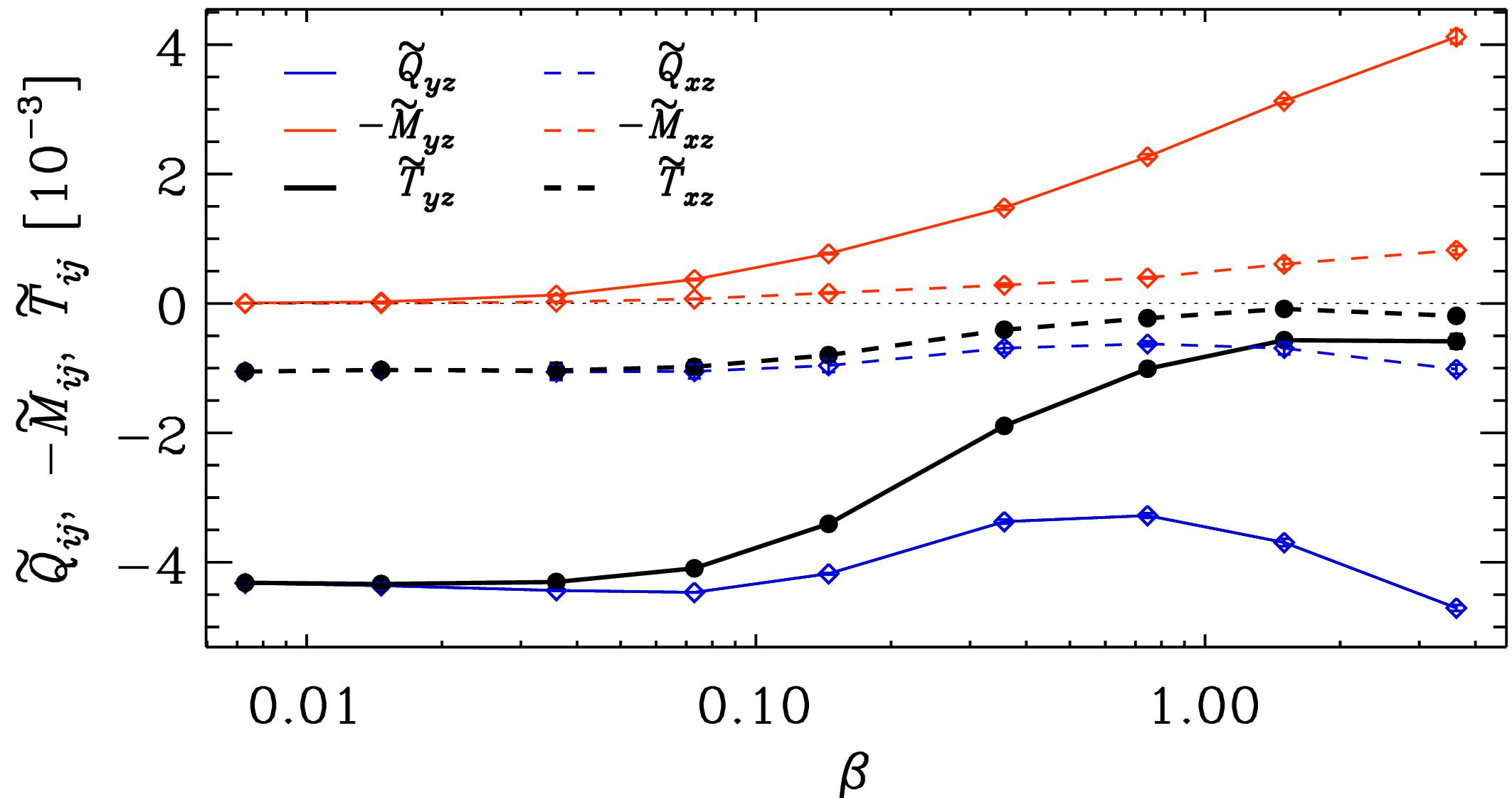}
\caption{Normalized vertical ($yz$, solid) and meridional ($xz$,
  dashed) Reynolds (blue), Maxwell (red), and total (thick black)
  stresses as functions of the normalized magnetic field strength
  $\beta$.  Data from $\theta=90\degr$ ($50\degr$) for the vertical
  (meridional) stress is shown.}
\label{fig:pqyzqxz_lsf}
\end{figure}

Significant magnetic quenching is apparent for all coefficients except
$H^{(0)}$ clearly before equipartition strength is reached.  The
behaviour of the $\Lambda$ coefficients is similar in the vertical and
horizontal field cases (compare the solid and dashed lines in
\Figa{fig:pLambda_lsf}). However, noticeable quenching occurs at
somewhat lower magnetic field strengths in the LSFV runs.  The
analytic and numerical results show qualitatively similar behaviour. The magnetic quenching occurs at somewhat lower magnetic fields
in the simulations in comparison to theory. For magnetic fields near
equipartition, the $\Lambda$ coefficients have diminished to roughly
10--20 \%\ of their hydrodynamic values. The absolute maximum
value for $H^{(0)}$ is obtained for $\Lu=10$ (Set~LSFH7) corresponding
to $B_0\approx0.8\Beq$, after which its magnitude also decreases. This
is somewhat lower than the analytically predicted value of
$\beta\approx0.94$. The apparently deviating behaviour, i.e.
increasing magnitude for $\beta\gtrsim 0.3$ of $V^{(1)}$, $H^{(2)}$,
and $M^{(0)}$, is due to the change in   latitude distribution of the
stress  as a function of $\mBBB$. The magnetic quenching comes
about because the Maxwell stress has a similar latitude distribution,
but is of opposite sign  to the Reynolds
stress. Moreover, the Maxwell stress increases monotonically as a
function of the imposed magnetic field (see
\Figa{fig:pqyzqxz_lsf}). However,  the Reynolds stress also increases
for $\beta\gtrsim1$. The tendency of the Reynolds and Maxwell
contributions to cancel is reminiscent of the behaviour of the total
turbulent stress in semi-global convection simulations where small-
and large-scale dynamos are simultaneously excited
\citep{2017A&A...599A...4K}.

\section{Conclusions}

The non-diffusive contribution to the Reynolds stress, or the
$\Lambda$ effect, from numerical simulations of homogeneous
anisotropically forced turbulence was found to agree with analytic
theory derived under the second-order correlation approximation. This
includes the scaling of the off-diagonal Reynolds stress for slow
rotation and the proportionality of the vertical $\Lambda$ on the
vertical turbulence anisotropy $A_{\rm V}$. Furthermore, the Reynolds
stress is proportional to the Reynolds number at low $\Rey$.  At more
rapid rotation ($\Omega_\star \gtrsim 1$) the numerical results
indicate more complex latitude dependences than predicted by
theory. This entails a higher than second power of $\sin\theta$
for adequate fits of the data. At rapid rotation
($\Omega_\star\gtrsim 3\ldots 5$ depending on the stress component), a
strong rotational quenching was observed. This quenching is predicted
by theory \citep[e.g.][]{KR93,KR05}, but occurs at more rapid rotation
in the simulations.

The bulk of the current results are restricted to low values of the
Reynolds number ($\Rey=6 \ldots 14$). Current results indicate that
the Reynolds stress and the deduced $\Lambda$ effect become
independent of $\Rey$ between $10<\Rey<20$ at slow rotation. At rapid
rotation the system develops large-scale flows in the form of vortices
that have a profound influence on the dynamics. Such vortices are
ubiquitous in rapidly rotating turbulent systems
\citep[e.g.][]{1998PhFl...10.2895Y,Chan03,2014PhFl...26i6605F} at
sufficiently rapid rotation and Reynolds numbers. Thus it would appear
to be logical to assume that large-scale vortices would dominate the
dynamics in rapidly rotating astrophysical objects where the Reynolds
numbers are much higher than in the current simulations. However, the
vortices are also known to turn into jets in systems with horizontal
aspect ratios unequal to one \citep{2017PhRvF...2k3503G} and to
promote large-scale dynamo action \citep{2018A&A...612A..97B}. Strong
magnetic fields, however, can also quench the vortices \citep{KMB13}.

A strong quenching of the $\Lambda$ effect was found in the case where
an imposed vertical or horizontal uniform magnetic field was
introduced into the system. This is manifested by a decreasing total
stress, which is due to the Reynolds and Maxwell contributions having
opposite signs. This is reminiscent of recent convection-driven dynamo
simulations where a strong quenching of differential rotation was
attributed to a magnetically quenched $\Lambda$ effect
\citep{2017A&A...599A...4K}. The current results also confirm the
analytic prediction \citep{KPR94} of an $H^{(0)}$ component which is
due to an additional anisotropy introduced by an imposed horizontal
field. Another remarkable aspect is that despite the low Reynolds
numbers of the current simulations, they are still well outside the
formal validity range of second-order correlation approximation
\citep[e.g.][]{KR80}, yet the numerical results are at least in
qualitative agreement with SOCA predictions.

The current results regarding the slow rotation regime and magnetic
quenching of the $\Lambda$ effect possibly open a window to the
estimation of the subsurface magnetic field in the Sun via the solar
cycle dependent near-surface shear
\citep{Ki16,2016A&A...595A...8B}. The missing piece of this puzzle is
the magnetic field dependence of the turbulent viscosity, that is
$\nut=\nut(\mBBB)$. Another pressing issue is the role of the
large-scale vortices in the angular momentum transport in the rapidly
rotating regime. Such studies are, however, likely to require more
sophisticated methods to extract the turbulent transport coefficients.

\begin{acknowledgements}
  We thank the anonymous referee for the constructive comments.
  Atefeh Barekat, Axel Brandenburg, Maarit K\"apyl\"a, Igor
  Rogachevskii, G\"unther R\"udiger, and J\"orn Warnecke are
  acknowledged for their valuable comments on the manuscript. The
  simulations were performed using the supercomputers hosted by CSC --
  IT Center for Science Ltd.\ in Espoo, Finland,  administered
  by the Finnish Ministry of Education. Financial support from the
  Academy of Finland Centre of Excellence ReSoLVE (Grant No.\ 307411)
  and the Deutsche Forschungsgemeinschaft Heisenberg programme (Grant
  No.\ KA 4825/1-1) is acknowledged.
\end{acknowledgements}

\bibliographystyle{aa}

\end{document}